\documentclass[journal,onecolumn]{IEEEtran}
\IEEEoverridecommandlockouts
\usepackage{amsmath,amsfonts}
\usepackage{amssymb}
\usepackage{algorithmic}
\usepackage{algorithm}
\usepackage{array}
\usepackage[caption=false,font=footnotesize,labelfont=sf,textfont=sf]{subfig}
\usepackage{textcomp}
\usepackage{stfloats}
\usepackage{url}
\usepackage{cite}
\usepackage{verbatim}
\usepackage{graphicx}
\usepackage{float}
\usepackage{subfig}
\usepackage{booktabs}
\usepackage{multirow}
\usepackage{color}
\usepackage{makecell}
\usepackage{soul}
\usepackage[colorlinks,citecolor=blue,linkcolor=blue]{hyperref}
\usepackage[numbers,sort&compress]{natbib}
\newtheorem{corollary}{Corollary}
\newtheorem{proposition}{Proposition}
\newtheorem{theorem}{Theorem}

\newtheorem{lemma}{Lemma}

\newtheorem{definition}{Definition}
\newtheorem{example}{Example}
\makeatletter

\newcommand{\Rmnum}[1]{\expandafter\@slowromancap\romannumeral #1@}
\makeatother

\def\BibTeX{{\rm B\kern-.05em{\sc i\kern-.025em b}\kern-.08em
    T\kern-.1667em\lower.7ex\hbox{E}\kern-.125emX}}
\begin{document}

\title{Channel Polarization under Channel Noise with Memory}

\author{Tianfu Qi, \emph{Graduate Student Member, IEEE}, Jun Wang, \emph{Senior Member, IEEE}
\thanks{Tianfu Qi, Jun Wang are with the National Key Laboratory on Wireless Communications, University of Electronic Science and Technology of China, Chengdu 611731, China (e-mail: 202311220634@std.uestc.edu.cn, junwang@uestc.edu.cn).}
}

\markboth{Qi \MakeLowercase{\textit{et al.}}: Channel Polarization under Channel Noise with Memory}%
{Shell \MakeLowercase{\textit{et al.}}: Bare Demo of IEEEtran.cls for IEEE Journals}

\maketitle

\begin{abstract}
The channel polarization behavior of polar codes under noise with memory is investigated. By introducing a genie-aided channel model, we first show that the polarized subchannels still converge to extremal channels under the standard polar coding framework. More importantly, we explicitly quantify the gap between the mutual information achieved by ignoring memory effects and the actual capacity attained after sufficient polarization. It is proven that the channel capacity remains achievable even without prior knowledge of the channel noise. Furthermore, we demonstrate that the polarization rate is slower than that in the binary-input memoryless channel (BMC) case, provided that the channel transition function satisfies certain conditions. In particular, the Bhattacharyya parameter is asymptotically upper-bounded and lower-bounded by a polynomial function and an exponential function with respect to the block length, respectively.
\end{abstract}


\section{Introduction}
\IEEEPARstart{A}{s} a breakthrough in coding theory, polar codes are the first class of codes proven to achieve the capacity of binary-input discrete memoryless channels (B-DMCs) \cite{paper1}. The fundamental principle underlying polar codes is channel polarization, which is realized through an iterative transformation of the input sequence using a polarization kernel, i.e., $F=\left(
\begin{array}{cc}
  1 & 0 \\
  1 & 1
\end{array}
\right)$. The generator matrix of size $L$ is constructed as the Kronecker product of $\log L$ polarization kernels\footnote{If not specified, the base of the logarithm is 2 by default in this paper.}, followed by a permutation operation. In this process, $L$ identical copies of the original channel are combined to form a synthesized channel, where $L$ represents the block length. This synthesized channel is then split into $L$ subchannels with varying levels of reliability. As $L$ approaches infinity, the reliability of these subchannels polarizes toward two extremal values: one corresponding to a perfect channel with a capacity equal to 1, and the other to a completely noisy channel that is unusable for reliable transmission. Accordingly, information bits are assigned to the highly reliable subchannels, while the unreliable ones are fixed with frozen bits, which are predetermined and shared between the transmitter and receiver. In the case of B-DMCs, the fraction of perfect subchannels among all $L$ synthesized channels asymptotically equals the capacity of the original channel, denoted by $I(W)$. Thus, polar coding enables efficient utilization of reliable subchannels to achieve transmission rates approaching the channel capacity.

The polarization phenomenon has been extensively studied under the assumption of memoryless channel noise. However, in many practical communication scenarios, noise exhibits memory. Examples include NOMA system \cite{paper5}, IoT wireless communication \cite{paper6}, powerline communication \cite{paper9}, LTE systems in urban environments, digital video broadcasting-terrestrial (DVB-T), etc. In such cases, noise samples are not mutually independent and often exhibit clustering behavior \cite{paper5, paper6, paper7, paper8, paper9, paper16}. Consequently, several assumptions underpinning the analysis in the memoryless setting no longer hold. For example, the mutual information (MI) relationship between the synthesized channel and the original channel is derived based on the assumption of channel independence, that is, the noise is memoryless. Moreover, the transformation of the Bhattacharyya parameter becomes significantly more complex in the presence of memory, as both the channel outputs and noise samples are correlated. This correlation makes it difficult to isolate the random variables involved in the summation (or integration, in the case of continuous output alphabets), thereby preventing the decomposition of the Bhattacharyya parameter of the combined channel into those of the split subchannels.

Şaşoğlu is among the first to investigate the polarization process in channels with memory, as presented in \cite{paper2}. He proved that polarization occurs for any $q$-ary input sequence governed by a $k$-th order Markov process, where $k$ is a positive integer and $q$ is a prime number. The main conclusions align with those originally proposed by Arikan. Subsequently, the analysis was extended to more general processes, specifically $\psi$-mixing processes, in \cite{paper3}, where the constraint of prime Markov order was also removed. It was further shown that the polarization rate remains the same as in the memoryless case. Shuval et al. investigated polar codes for the processes modeled by finite-state aperiodic irreducible Markov (FAIM) chains \cite{paper4}, which generalize conventional Markov processes. Their approach involves partitioning the FAIM process into blocks and leveraging state dependencies between adjacent blocks to establish fast polarization under memory. A similar and more refined analysis is also provided in \cite{paper12}. These works prove that the polar coding techniques are still efficient in presence of memory.

In this paper, we focus on channels where the noise exhibits memory\footnote{Throughout the paper, we assume that the input bits are mutually independent. Under this assumption, the terms `channel with memory' and `noise with memory' are used interchangeably.}. Many practical channels fall into this category, including the Gilbert-Elliott (GE) channel \cite{paper18,paper19,paper20}, Markov-Middleton (MM) noise model \cite{paper21}, Bernoulli-Gaussian (BG) channel \cite{paper22}, $\alpha$ sub-Gaussian ($\alpha$SG) noise model \cite{paper8,paper24}, and Gaussian-Student (GS) noise model \cite{paper16}, among others. In this work, however, we consider a general noise model with memory. \textit{In addition to investigating channel polarization under noise with memory, we also address a more practical issue.} In many scenarios, detection or decoding algorithms are designed without accounting for memory effects, often due to practical constraints. For instance, the channel state information (CSI) may vary rapidly over time, making its estimation challenging and computationally expensive. As a result, the memory characteristics of the noise may become difficult to detect, increasing the likelihood of overlooking temporal correlations. This ignorance can lead to significant performance degradation. Coding techniques can help mitigate this degradation, as they inherently introduce correlations among transmitted symbols. Given the structured nature and theoretical appeal of polar codes, our goal is to quantify their potential benefits in the presence of noise with memory.

Motivated by the above considerations, the main contributions of this work are summarized as follows:
\begin{itemize}
\item{\textbf{Theoretical aspects:} In contrast to the approaches in \cite{paper2,paper3,paper4}, we present a new proof for the convergence of the MI process $\{I_n\}$ under channel noise with memory. To address the analytical challenges posed by memory effects, we introduce a genie-aided (GA) channel. Although the GA channel is physically unrealizable since it assumes perfect knowledge of all prior noise samples, it serves as a powerful analytical tool to facilitate the derivation. We then prove that the average mutual information after channel polarization equals the achievable capacity, by analyzing the convergence behavior of the Bhattacharyya process $\{Z_n\}$. Moreover, we observe that the polarization rate is substantially slower than that in the memoryless case. Specifically, while the polarization rate for binary-input memoryless channels (BMCs) is exponential in the square root of the block length, it becomes only polynomial when the noise exhibits memory, provided certain conditions on the noise model are satisfied. \textit{It is important to emphasize that, although our proof is conceptually inspired by Arikan's martingale-based framework, it cannot be directly applied in the presence of memory. Several non-trivial modifications are required to accommodate the temporal correlations introduced by the noise process.}}
\item{\textbf{Practical insights:} We show that the average capacity of all polarized subchannels is greater than or equal to the MI of the original channel if the memory property is ignored. The performance gap depends explicitly on the channel transition function. In particular, for channels with the noise following a first-order Markov process, we prove that the gap $c=I(N_1;N_0)-\frac{1}{2}I(Y_1;Y_0)-I(Y_0;Y_1^{+\infty})\geq0$ where $Y_j$ and $N_j$ denote the channel output and noise random variables (RVs) at time $j$, respectively. This result implies that the performance loss caused by neglecting noise memory can be asymptotically eliminated as the block length increases, even without prior knowledge of the noise process. Therefore, in scenarios where the channel noise exhibits memory, prior information about the noise may become less critical when polar coding is employed. Finally, as a complement to the main results, we also discuss the extensions of our analytical framework and potential research and application directions.}
\end{itemize}

The remainder of this paper is organized as follows. Section \ref{PROBLEM_STATEMENT} presents the problem formulation and parameter definitions along with the key research questions addressed in this work. Section \ref{PROOF_OF_THEOREM_1} and \ref{PROOF_OF_THEOREM_4_5} provide the proofs concerning the transformations of mutual information and the Bhattacharyya parameter processes, respectively. Extensions of the main results to more general settings and future research directions are discussed in section \ref{EXPANSION_AND_DISCUSSION}. Finally, the whole paper is concluded in section \ref{CONCLUSION}.

\emph{Notations:} We utilize uppercase and lowercase to represent the RV and its realizations, i.e., $X$ and $x$. The random vector is denoted by $\vec{X}$. $x_a^b$ denotes the vector $[x_a,\cdots,x_b]$ and $x_a^b={\emptyset}$ if $a>b$. $E_X[\cdot]$ denotes the expectation operation with respect to $X$. The $I(\cdot)$, $Z(\cdot)$ separately represent the mutual information and Bhattacharyya parameter function. `$\oplus$' denotes the modular-2 operation. The $\chi_i^j=\{i,\cdots,j\}$ denotes the indicator set.

\section{Preliminaries}\label{PROBLEM_STATEMENT}
In this section, we begin with a brief review of polar codes and introduce several definitions that will be used in what follows. We then outline the central problems addressed in this work.

\subsection{Polar codes and settings}
\begin{figure}[htbp]
\centering
\includegraphics[width=5in]{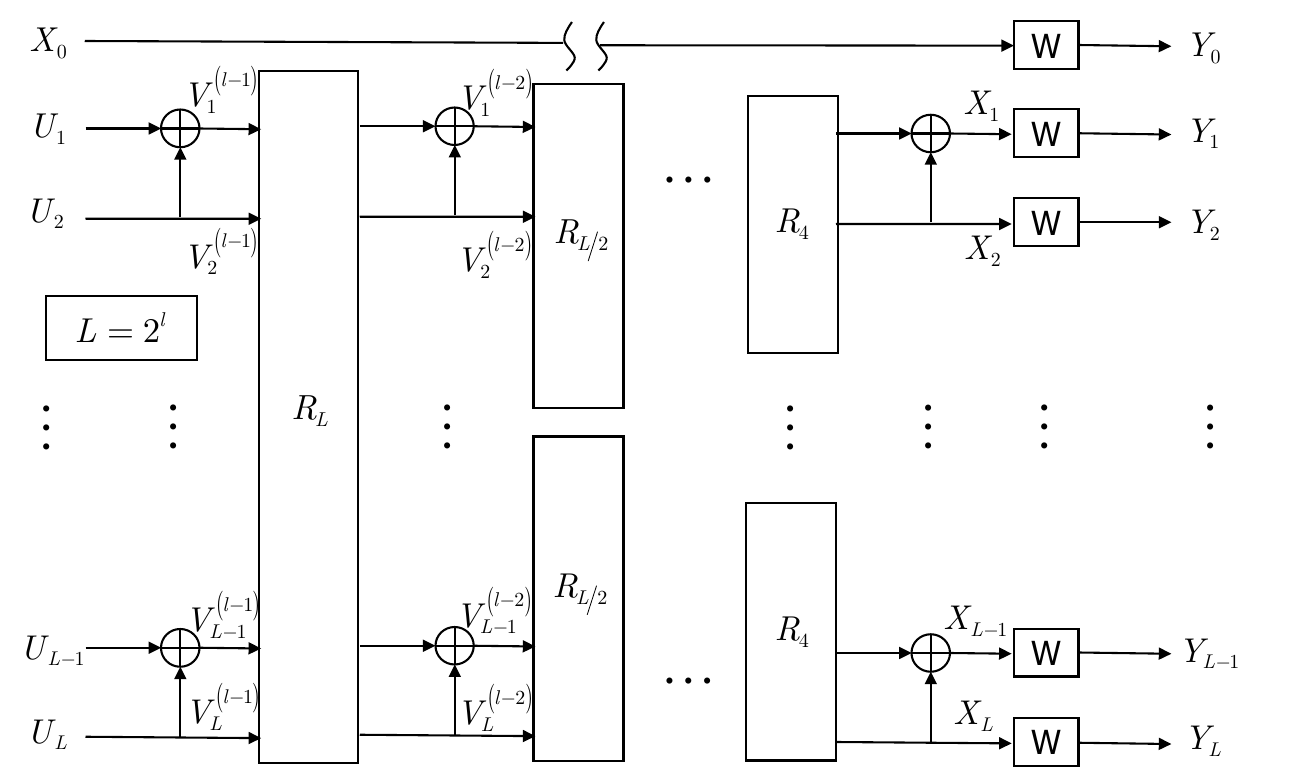}\label{polarization_process}
\caption{The framework of the standard polar encoder with block length $L=2^l$.}
\label{fig_1}
\end{figure}

The polarization framework and the notations used in this paper are illustrated in Fig. \ref{fig_1}. The block length is given by $L = 2^l$. Let $\{U_i\}$ and $\{X_i\}$ denote the source information and the coded information random processes, respectively, where $i$ represents the time index. The inner bits are denoted as $V_j^{(k)}$, where $j$ and $k$ correspond to the number of subchannels and layers, respectively. The polarization structure for $L = 2$ is referred to as the polarization unit. Additionally, $Y_0$ represents the final output of the previous block, while $R_L$ denotes the permutation operation applied to $L$ inputs.

Without loss of generality, we assume that the noise process $\{N_i\}$ exhibits memory and follows a first-order Markov process, meaning that each noise sample is only dependent on the immediately preceding sample. Higher-order and fractional-order processes can be analyzed similarly. Let the channel function be denoted by $W: \mathcal{X} \rightarrow \mathcal{Y}$. Our primary focus is on binary input, so we set $\mathcal{X} = \mathcal{B} = \{0, 1\}$. The channel output is assumed to be continuous-valued, and the mutual information $I(W)$ and Bhattacharyya parameter $Z(W)$ of the channel $W$ are defined as $I(W)=\int_{\mathcal{R}}\sum_{X\in\mathcal{B}}W(y,x)\log\frac{W(y|x)}{W(y)}dy$ and $Z(W)=\int_{\mathcal{R}}\sqrt{W(y|0)W(y|1)}dy$. Other parameters related to the conditional probability density function (PDF) of the channel can be similarly defined.

Next, we provide the definition of the GA channel.
\begin{definition}\label{definition_1}
\textit{Let the channel noise $\{N_i\}$ be modeled as a $p$-th order Markov process. Then, the GA channel is defined as
\begin{equation}
\tilde{W}: \mathcal{X} \times \mathcal{N}^{p} \rightarrow \mathcal{Y}
\end{equation}
where $\tilde{W}$ is assumed to have access to previous $p$ noise samples. Specifically, for the $i$-th time instant, the channel is represented by $\tilde{W}(y_i | x_i, n_{i-p}^{i-1})$.}
\end{definition}

\begin{figure}[htbp]
\centering
\includegraphics[width=4in]{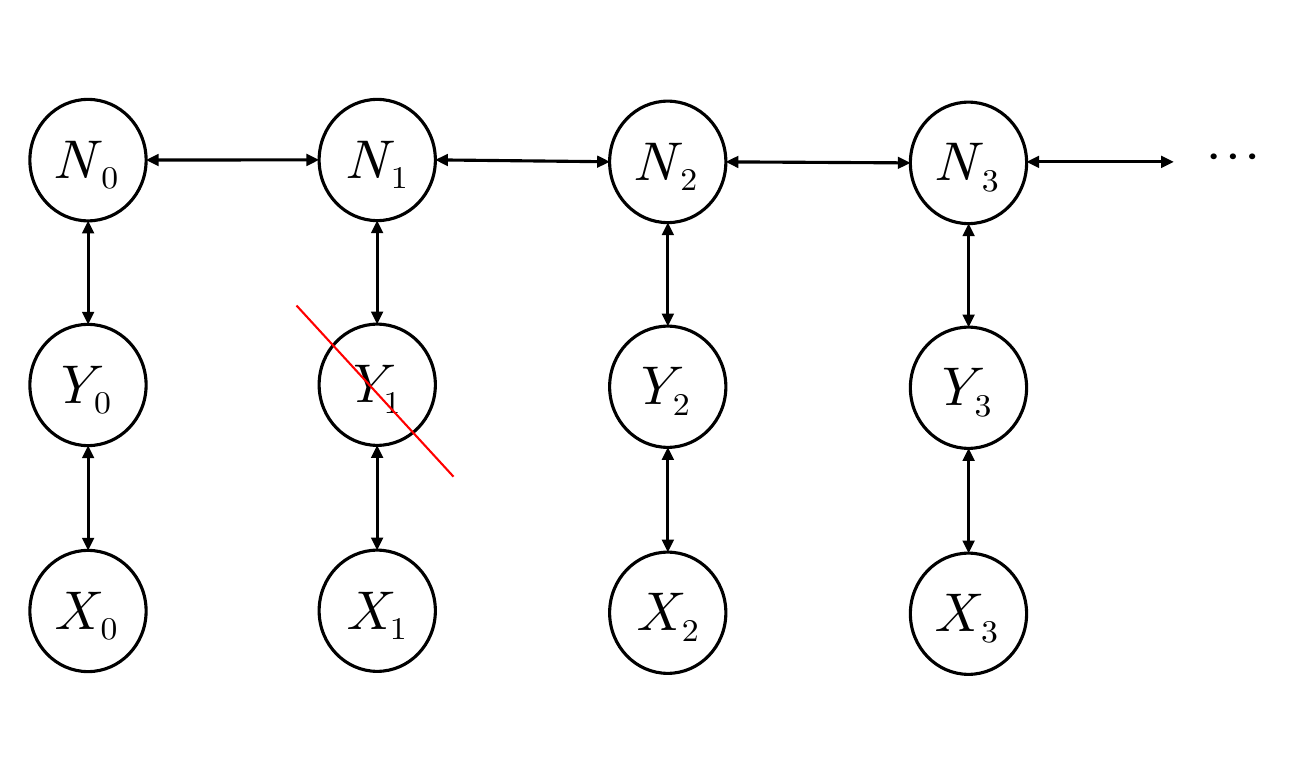}
\caption{Fig. \ref{Markov_process_example} illustrates the first-order Markov process, capturing the dependencies among the channel input $X$, channel noise $N$ and channel output $Y$. In this graph, two variables are considered dependent if there exists a connecting path between their corresponding vertices. For instance, even if the node corresponding to `$Y_1$' is removed, `$Y_0$' and `$Y_2$' remain connected, indicating that $W(y_{2}|y_{0})\neq W(y_{2})$. This demonstrates that the channel outputs exhibit temporal correlation beyond adjacent time steps.}
\label{Markov_process_example}
\end{figure}

For the first-order Markov noise process, $\tilde{W}$ simplifies to $\tilde{W}: \mathcal{X} \times \mathcal{N} \rightarrow \mathcal{Y}$ due to $\tilde{W}(y_i | x_i, n_{i-p}^{i-1}) = \tilde{W}(y_i | x_i, n_{i-1})$ with $p=1$, which corresponds to the noise PDF. Clearly, $\tilde{W}$ is not available, as the $n_{i-1}$ is unknown in practice. However, we demonstrate that $\tilde{W}$ is a useful theoretical tool for deriving the polarization of channel noise with memory. It is worth noting that $\tilde{W}$ can also be equivalently expressed as $\tilde{W}: \mathcal{X}^2 \times \mathcal{Y} \rightarrow \mathcal{Y}$, given that $\tilde{W}(y_i | x_i, n_{i-1}) = \tilde{W}(y_i | x_i, x_{i-1}, y_{i-1})$. It is important to emphasize that the channel output sequence does not constitute a first-order Markov process, that is, the conditional distribution does not satisfy $W(y_{i+1}|y_{i},y_{i-1})\neq W(y_{i+1}|y_{i})$. With slight abuse of notation, we use $W(y_{i+1}|y_i)$ to denote the conditional PDF induced by the transition function $W(\cdot|\cdot)$. In fact, the overall process forms a more complex dependency structure, which can be illustrated in Fig. \ref{Markov_process_example}. As shown, the $i$-th output sample depends on all preceding outputs, indicating long-range temporal correlation.

In \cite{paper1}, polarization is achieved through channel combining and splitting, which can also be applied to the GA channel. Specifically, we have $W_L: \mathcal{X}^L \rightarrow \mathcal{Y}^L$, $\tilde{W}_L: \mathcal{X}^{L+1} \times \mathcal{Y} \rightarrow \mathcal{Y}^L$ and
\begin{align}\label{channel_combining}
\tilde{W}_L(y_1^L|x_1^L,x_0,y_0)=\prod_{i=1}^{L}\tilde{W}(y_i|x_i,x_{i-1},y_{i-1}),
\end{align}
where $x_0$ and $y_0$ represent virtual input and output arguments. \eqref{channel_combining} corresponds to the channel combining operation, and the channel splitting is
\begin{align}\label{channel_splitting}
\tilde{W}_L^{(i)}(y_1^L,&u_1^{i-1}|u_i,x_0,y_0)=\frac{1}{2^{L-1}}\sum_{U_{i+1}^L\in\mathcal{B}^{L-i}}\tilde{W}_L(y_1^L|x_1^L,x_0,y_0).
\end{align}

Note that \eqref{channel_combining} and \eqref{channel_splitting} are formulated for $\tilde{W}$ and the corresponding version of $W$ can be obtained directly by
\begin{equation}
\mathbb{E}_{N_{i-1}}[\tilde{W}(y_i|x_i, n_{i-1})] = W(y_i|x_i).
\end{equation}

For clarity, we simplify the notation as $\mathbb{E}_{N_{i-1}}[\tilde{W}] = W$ when no confusion arises in what follows. Then, we define the conditional Bhattacharyya parameter for GA channels.
\begin{definition}\label{definition_2}
\textit{Let the channel noise $\{N_i\}$ be modeled as a first-order Markov process. The conditional Bhattacharyya parameter for the GA channel is then defined as
\begin{equation}\label{definition_Z_g_n0}
Z_g(\tilde{W}|n_0)\triangleq\int_{\mathcal{R}}\sqrt{\tilde{W}(y_1|0,n_0)\tilde{W}(y_1|1,n_0)}dy_1,
\end{equation}
and
\begin{equation}\label{definition_Z_g}
Z_g(W)\triangleq\mathbb{E}_{N_0}[Z_g(\tilde{W}|n_0)].
\end{equation}}
\end{definition}

It is worth noting that $Z_g(W)$ reduces to the standard $Z(W)$ when the channel noise is memoryless. Moreover, by applying Jensen's inequality, it can be readily shown that
\begin{equation}
Z_g(W)\leq Z(W),
\end{equation}
which is an important property for the convergence of the sequence $\{Z_n\}$.


\subsection{Main problems}
The key problem addressed in this work is whether the conclusions established for memoryless channels remain valid in the presence of memory, and to what extent polar coding can provide performance gains compared to scenarios where no prior information is exploited. Specifically, we investigate whether the following relations hold,
\begin{equation}
2I(W_L^{(i)})=I(W_{2L}^{(2i-1)})+I(W_{2L}^{(2i)}),
\end{equation}
\begin{equation}
I(W_{2L}^{(2i-1)})\leq I(W_{L}^{(i)})\leq I(W_{2L}^{(2i)}).
\end{equation}

Moreover, it is reasonable to conjecture that
\begin{equation}
\lim\limits_{L\rightarrow+\infty}\frac{1}{L}I(W_L^{(i)})=I(W)+c,
\end{equation}
where $c\geq0$ is a constant, as coding typically yields noticeable performance gains. Nevertheless, our goal is to quantitatively characterize the value of $c$. To apply Arikan's martingale framework to derive the average capacity of polarized subchannels, it is essential to demonstrate the convergence of the sequence $\{Z_n\}$. However, the proof requires significant modifications due to the presence of noise memory.

Finally, the polarization rate must be discussed, which is equivalent to investigating the convergence rate of $\{Z_n\}$. In the memoryless case, it has been proven that \cite{paper10}
\begin{equation}
\lim\limits_{L\rightarrow+\infty}\Pr[Z_L^{(i)}\leq2^{-2^{l\beta}}]=1,
\end{equation}
for any $i\in\mathcal{A}_L(W)$ and $\beta<0.5$ where $\mathcal{A}_L(W)$ denotes the information bit index set. However, when the channel noise samples exhibit correlation, the burst errors will be occur more frequently during transmission. Therefore, it is reasonable to expect that the detection error upper bound will be larger compared to that in the memoryless channel for the same block length. In other words, the polarization process may occur more slowly than that in the memoryless case. In the following, we will demonstrate that only some of the above conclusions retain the same as the memoryless case.

\section{Convergence of $\{I_n\}$}\label{PROOF_OF_THEOREM_1}
In this section, we begin by deriving the MI transformation for the simplest case, namely $L=2$, and then extend the result to arbitrary block lengths.

The recursive channel transformation under noise with memory is first presented without proof. In fact, the procedure closely follows the framework in \cite{paper1} and can be verified straightforwardly.
\begin{proposition}\label{proposition_1}
\textit{Let $\tilde{W}_L^{(i)}$ be the polarized subchannel defined in \eqref{channel_splitting}. Then,
\begin{align}\label{proposition_1_equation_1}
\tilde{W}_{2L}^{(2i-1)}(y_1^{2L},u_1^{2i-2}|u_{2i-1},n_0)=\frac{1}{2}\sum_{U_{2i}}\tilde{W}_{L}^{(i)}(y_{L+1}^{2L},u_{1,e}^{2i-2}|u_{2i},n_L)\tilde{W}_{2L}^{(2i-1)}(y_1^{2L},u_{1,e}^{2i-2}\oplus u_{1,e}^{2i-2}|u_{2i-1}\oplus u_{2i},n_0),
\end{align}
\begin{align}\label{proposition_1_equation_2}
\tilde{W}_{2L}^{(2i)}(y_1^{2L},u_1^{2i-1}|u_{2i},n_0)=\frac{1}{2}\tilde{W}_{n}^{(i)}(y_{L+1}^{2L},u_{1,e}^{2i-2}|u_{2i},n_L)\tilde{W}_{2L}^{(2i-1)}(y_1^{2L},u_{1,e}^{2i-2}\oplus u_{1,e}^{2i-2}|u_{2i-1}\oplus u_{2i},n_0).
\end{align}}
\end{proposition}

Proposition \ref{proposition_1} confirms the feasibility of implementing the recursive channel transformation under noise with memory\footnote{Unless otherwise specified or required for clarity, the feasible sets of $U_{2i}$ in \eqref{proposition_1_equation_1} and other similar arguments are omitted. Throughout this paper, we assume by default that $U_1^p\in\mathcal{B}^p$.}. The following lemma further considers the MI transformation for $L=2$.
\begin{lemma}\label{lemma_1}
\textit{Let the code length $L$ be 2. Then,
\begin{align}\label{lemma_1_equation}
I(W_2^{(1)})+I(W_2^{(2)})&=2I(W)+I(N_1;N_0)-I(Y_1;Y_0)\nonumber\\
&\geq2I(W).
\end{align}}
\end{lemma}

\begin{IEEEproof}
Based on the definition of the GA channel $\tilde{W}(y_1|x_1,n_0)$, its mutual information is given by $I(Y_1;X_1,N_0)$. However, the channel capacity cannot be characterized by $I(Y_1;X_1,N_0)$, since $N_0$ is an auxiliary variable that does not contribute to the transmission rate. In other words, although $N_0$ can be utilized in decoding and detection, it can be considered as the state information and is not part of the information to be recovered. Specifically, we have
\begin{align}\label{relation_of_IW_and_IW}
I(Y_1;X_1,N_0)=&I(Y_1;X_1)+I(Y_1;N_0|X_1)\nonumber\\
=&I(Y_1;X_1)+I(N_1;N_0).
\end{align}

Consider a highly noisy channel in which the receiver obtains negligible information from the received signal. In this case, we have $I(Y_1;X_1)\approx 0$ while $I(N_1;N_0) > 0$, indicating that the mutual information of the GA channel remains strictly positive even when the transmitted signal power approaches zero. Clearly, this result lacks physical meaning, and further confirms that $\tilde{W}$ is not an achievable channel model. Nevertheless, the relation given in \eqref{relation_of_IW_and_IW} can be exploited to compute $I(Y_1;X_1)$ for further discussion. We first consider the sum of polarized GA subchannels as follows,
\begin{align}\label{lemma_1_equation_1}
&I(\tilde{W}_2^{(1)})+I(\tilde{W}_2^{(2)})\nonumber\\
=&I(Y_1^2;U_1,N_0)+I(Y_1^2,U_1;U_2,N_0)\nonumber\\
\overset{(a)}{=}&I(Y_1^2;U_1,N_0)+I(Y_1^2;U_2|N_0,U_1)+I(Y_1^2;N_0|U_1)\nonumber\\
\overset{(b)}{=}&I(Y_1^2;X_1^2,N_0)+I(N_0;Y_2|Y_1,U_1)+I(Y_1;N_0|U_1)\nonumber\\
=&I(Y_1^2;X_1^2,N_0)+I(Y_1^2,U_1;N_0),
\end{align}
where $(a)$ follows the chain rule, and $(b)$ holds because there exists a bijection between $(U_1, U_2)$ and $(X_1, X_2)$. For the case of memoryless channels, the mutual information $I(Y_1^2; X_1^2)$ decomposes as $2I(Y; X)$ based on the independence of the channel outputs. However, this decomposition does not hold when the channel noise exhibits memory. Therefore, we instead partition $I(Y_1^2; X_1^2, N_0)$ to obtain the right-hand side (RHS) of \eqref{lemma_1_equation}, i.e.,
\begin{align}\label{lemma_1_equation_2}
&I(Y_1^2;X_1^2,N_0)\nonumber\\
=&I(Y_1;X_1,N_0)+I(Y_1;X_2|X_1,N_0)+I(Y_2;X_1^2,N_0|Y_1)\nonumber\\
\overset{(a)}{=}&I(Y_1;X_1,N_0)+I(Y_2;X_1^2,N_0,Y_1)-I(Y_2;Y_1)\nonumber\\
=&I(Y_1;X_1,N_0)+I(Y_2;X_1^2,Y_1)+I(Y_2;N_0|X_1^2,Y_1)-I(Y_2;Y_1)\nonumber\\
\overset{(b)}{=}&2I(\tilde{W})-I(Y_2;Y_1),
\end{align}
where $(a)$ follows the fact that for the first-order Markov process, $X_2$ provides no additional information about $Y_1$ given $X_1$ and $N_0$. A similar reasoning applies to $(b)$, where $I(Y_2; N_0 | X_1^2, Y_1) = 0$ holds. By combining \eqref{lemma_1_equation_1} and \eqref{lemma_1_equation_2},
\begin{align}
&I(\tilde{W}_2^{(1)})+I(\tilde{W}_2^{(2)})-2I(\tilde{W})\nonumber\\
=&I(Y_1^2,U_1;N_0)-I(Y_2;Y_1)\nonumber\\
\overset{(a)}{=}&I(Y_1;X_0,Y_0)+I(Y_2,U_1;X_0,Y_0|Y_1)-I(Y_2;Y_1)\nonumber\\
\overset{(b)}{=}&I(Y_1;X_0|Y_0)+I(Y_2,U_1;X_0,Y_0|Y_1)\geq0,
\end{align}
where $(a)$ follows $I(Y_1^2, U_1; N_0) = I(Y_1^2, U_1; X_0, Y_0)$ and $(b)$ stems from $I(Y_2; Y_1) = I(Y_1; Y_0)$. Finally, by applying \eqref{relation_of_IW_and_IW} and performing some manipulations, lemma \ref{lemma_1} is established.
\end{IEEEproof}

The lemma \ref{lemma_1} indicates that the total mutual information of two original channels increases after one layer of polarization.
\textbf{In the following, we provide a detailed explanation of how this increase in mutual information arises, which can be interpreted from two complementary perspectives.}

\begin{enumerate}
\item{From the perspective of detection theory, when considering two symbols transmitted independently through the original channel without additional prior information, the decision statistics are computed based on $W(y_1|x_1)$ and $W(y_2|x_2)$, respectively. This approach disregards the correlation between noise samples, leading to a performance loss, and the resulting mutual information is limited to $2I(W)$. In contrast, after polarization, the two channels are jointly considered, which enables more effective exploitation of the noise correlation, thereby improving performance.

    In particular, for $I(Y_1;X_1)$, we determine the $X_1$ given $Y_1$ by $W(y_1|x_1)$, which is derived from
     \begin{equation}
     W(y_1|x_1)=\int_{N_0\in\mathcal{R}}W(y_1|x_1,n_0)f_{N_0}(n_0)dn_0,
     \end{equation}
     or equivalently,
     \begin{equation}
     W(y_1|x_1)=\int_{Y_0\in\mathcal{R}}W(y_1|x_1,y_0)f_{Y_0}(y_0)dy_0,
     \end{equation}
     where $f_{Y_0}(\cdot)$ and $f_{N_0}(\cdot)$ represent the PDFs of $Y_0$ and $N_0$, respectively. This marginalization process disregards prior information such as $n_0$ or $y_0$, which makes $W(y_1|x_1)$ a less accurate representation compared to the conditional distributions $W(y_1|x_1,n_0)$ and $W(y_1|x_1,y_0)$.

    In contrast, for $I(Y_1^2;U_1)$, the polarized transition probability takes the form $W(y_1^2|u_1) = W(y_2|u_1,y_1) W(y_1|u_1)$. In the case of memoryless noise, the underlying process of $W(y_2|u_1,y_1)$ can be interpreted as follows: given $y_1$, we estimate $\hat{X}_1$, which then yields an estimate $\hat{X}_2$ (or equivalently $\hat{U}_2$) combined with prior information $U_1$. Then, $Y_2$ is generated via $W(y_2|\hat{x}_2)$. However, for noise with memory, the generation of $Y_2$ involves the transition function $W(y_2|\hat{x}_2,y_1)$ instead of $W(y_2|\hat{x}_2)$. Due to the inherent correlation between noise samples, $Y_1$ not only contributes to estimating $X_1$, but also plays a role in refining the effective noise model used to generate $Y_2$. In other words, after one-layer polarization, the underlying channel representation becomes more accurate in capturing the statistical characteristics of the channel noise. This phenomenon does not occur in the case of memoryless noise.}
\item{In terms of the derivation process, it can be observed that the MI difference between a channel and its corresponding GA version is given by $I(N_1;N_0)$, which we refer to as noise correlation information (NCI) for simplicity. Compared to $2I(W)$, the MI of the GA channel $2I(\tilde{W})$ increases by an amount equal to 2 NCI. However, for the polarized subchannel, we have $I(\tilde{W}_2^{(1)}) + I(\tilde{W}_2^{(2)}) = I(Y_1^2; X_1^2, N_0)$, which exhibits only a single NCI increase relative to $I(Y_1^2; X_1^2)$. This is because part of the NCI has already been partially utilized in constructing $I(Y_1^2; X_1^2)$. Here, `partially' means that the MI gain is smaller than the full NCI, specifically quantified as $I(N_1; N_0) - I(Y_1; Y_0)$, which corresponds to the result presented in Lemma \ref{lemma_1}.}
\end{enumerate}

Before proceedings, we need to introduce two additional lemmas.

\begin{lemma}\label{lemma_2}
\textit{Let $Y$ be the output of the channel with memory noise. Then, the sequence $\{I(Y_0;Y_i|Y_1^{i-1})\}$ is non-increasing and $\lim\limits_{i\rightarrow+\infty}I(Y_0;Y_i|Y_1^{i-1})=0$. Moreover, $I(Y_0;Y_1^{+\infty})<+\infty$.}
\end{lemma}
\begin{IEEEproof}
On one hand, $I(Y_0;Y_1^L) = I(Y_0;X_1,Y_1^L) - I(Y_0;X_1|Y_1^L)$. It is straightforward to demonstrate that $f(y_0|x_1,y_1,y_2^L) = f(y_0|x_1,y_1)$ using the Markov property. Consequently, $I(Y_0;X_1,Y_1^L) = I(Y_0;N_1) < +\infty$, which also holds as $L \to +\infty$. Therefore, it follows that $I(Y_0;Y_1^{+\infty}) < +\infty$. By applying the chain rule, we can have
\begin{align}\label{lemma_2_1}
I(Y_0;Y_1^L)=\sum_{i=1}^{L}I(Y_0;Y_i|Y_1^{i-1}).
\end{align}

Note that $I(Y_0;Y_{i+1}|Y_1^{i})=I(Y_0;Y_i|Y_1^{i-1},Y_{i+1})+I(Y_0;Y_{i+1}|Y_1^{i-1})-I(Y_0;Y_{i}|Y_1^{i-1})$. According to the data process inequality, it is easy to check $I(Y_0;Y_i|Y_1^{i-1},Y_{i+1})\leq I(Y_0;Y_i|Y_1^{i-1})$ and $I(Y_0;Y_{i+1}|Y_1^{i-1})\leq I(Y_0;Y_{i}|Y_1^{i-1})$. Therefore, we conclude that $\{I(Y_0;Y_i|Y_1^{i-1})\}$ forms a monotonic sequence. Combining equation \eqref{lemma_2_1} and the monotone convergence theorem, lemma \ref{lemma_2} is proved.
\end{IEEEproof}

\textbf{Remark 1:} Lemma \ref{lemma_2} aligns with intuitive expectations. For example, the correlation between two signal samples $Y_i$ and $Y_j$ weakens as the time interval $|i - j|$ increases. Suppose we aim to determine $Y_0$ given the prior information $Y_1^{L-1}$. In this case, the additional information provided by $Y_L$ diminishes as $L$ increases. It is important to note, however, that Lemma \ref{lemma_2} does not imply that $Y_0$ and $Y_L$ become independent as $L \to +\infty$, which is incorrect, as illustrated in Fig. \ref{Markov_process_example}.

\begin{lemma}\label{lemma_3}
\textit{Let the code length be $L=2^l$. Then,
\begin{align}\label{lemma_3_aim}
\sum_{i=1}^{l}2^{i-1}I\left(Y_1^{2^{l-i}};Y_{2^{l-i}+1}^{2^{l-i+1}}\right)=\sum_{i=1}^{L-1}I(Y_0;Y_1^{i}).
\end{align}}
\end{lemma}
\begin{IEEEproof}
We first convert the $I(Y_1^L;Y_{L+1}^{2L})$ to the mutual information between the single RV and random vector. For example,
\begin{align}
&I(Y_1^L;Y_{L+1}^{2L})\nonumber\\
=&I(Y_{L+1}^{2L};Y_{L})+I(Y_1^{L-1};Y_{L+1}^{2L}|Y_L)\nonumber\\
=&I(Y_{L+1}^{2L};Y_{L})+I(Y_1^{L-1};Y_{L}^{2L})-I(Y_1^{L-1};Y_L).
\end{align}

With similar procedures, it can be obtained that
\begin{align}\label{lemma_3_1}
I(Y_1^L;Y_{L+1}^{2L})=\sum_{i=L}^{2L-1}I(Y_0;Y_1^i)-\sum_{i=1}^{L-1}I(Y_0;Y_1^i).
\end{align}

By substituting \eqref{lemma_3_1} into \eqref{lemma_3_aim} and performing some manipulations, this lemma can be concluded.
\end{IEEEproof}

\begin{corollary}\label{corollary_4}
\textit{The sequence $\{I(Y_0;Y_1^{i})\}$ is increasing in terms of $i$ and bounded as follows,
\begin{equation}
\lim\limits_{i\rightarrow+\infty}I(Y_0;Y_1^{i})<I(Y_0;N_1)<+\infty.
\end{equation}
}
\end{corollary}

The corollary \ref{corollary_4} can be obtained directly from lemma \ref{lemma_2}. Now, we are ready to combine the above analysis and provide the following theorem.
\begin{theorem}\label{theorem_1}
\textit{With the standard polar coding structure and channels defined in \eqref{channel_combining} and \eqref{channel_splitting}, we have
\begin{align}
\lim\limits_{L\rightarrow+\infty}\frac{1}{L}\sum_{i=1}^{L}I(W_L^{(i)})=I(W)+I^{\dagger}(W),
\end{align}
where
\begin{align}
I^{\dagger}(W)=I(N_1;N_0)-\frac{1}{2}I(Y_1;Y_0)-I(Y_0;Y_1^{+\infty})\geq0.
\end{align}}
\end{theorem}
\begin{IEEEproof}
The proof is relegated to appendix \ref{appendix_1}.
\end{IEEEproof}

\textbf{Remark 2:} Theorem \ref{theorem_1} implies that the correlation between adjacent noise samples is fully exploited by polar coding as the block length becomes sufficiently large. When using the successive cancellation (SC) decoder, the average capacity of the polarized subchannels can exceed the MI of the original channel without memory consideration. Moreover, it is worth noting that the expression of $I(W_L^{(i)})$ does not involve any prior noise information, indicating that the performance loss caused by ignoring memory can be completely mitigated by polar coding as $L \rightarrow +\infty$.

\textbf{Remark 3:} It is important to emphasize that polar coding does not increase the channel capacity, even as $L \rightarrow +\infty$. This fact can be formally verified as follows,
\begin{equation}
C\triangleq\lim\limits_{L\rightarrow+\infty}\frac{1}{L}I(Y_1^L;X_1^L)=\lim\limits_{L\rightarrow+\infty}\frac{1}{L}I(Y_1^L;U_1^L)=\lim\limits_{L\rightarrow+\infty}\frac{1}{L}\sum_{i=1}^{L}I(W_L^{(i)}).
\end{equation}

\textbf{Remark 4:} We consider the memoryless noise channel as a special case to test the generality of our theorem. In this case, the output samples of the channel are mutually independent. Consequently, we have $I(N_0; N_1) = 0$ and $I(Y_0; Y_1) = 0$. Thus, it follows that $I^{\dagger}(W) = 0$ and the theorem \ref{theorem_1} reduces to the result in \cite{paper1}.

\textbf{Remark 5:} The relationship between the mutual information of adjacent split channels, i.e., $I(W_L^{(2i-1)})$ and $I(W_L^{(2i)})$, must be determined. In fact, the proof is quite straightforward. For example, $I(W_{L}^{(2i)})-I(W_{L}^{(2i-1)})=I(Y_1^L;U_{2i}|U_1^{2i-1})-I(Y_1^L;U_{2i-1}|U_1^{2i-2})=I(Y_1^L;U_1^{2i-2}|U_{2i-1}^{2i})\geq0$.

Actually, the relation can be explained from another aspect. First, we have

\begin{equation}
I(W_{L}^{(2i)})=I(Y_1^L,U_1^{2i-2};U_{2i})+I(U_{2i-1};U_{2i}|Y_1^L,U_1^{2i-2}).
\end{equation}

For the output of channel $Y_i$, it can be expressed as $Y_i=X_{i}^{\mathbf{b}_i}+N_i$ where $\mathbf{b}_i$ denotes the bit index used to generate $X_{i}$. For instance, for $L = 4$, we have $\mathbf{b}_4 = \{1, 2, 3, 4\}$ and $\mathbf{b}_2 = \{3, 4\}$ based on the generator matrix. Thus,
\begin{align}
&I(Y_1^L,U_1^{2i-2};U_{2i})\nonumber\\
=&I((X_{1}^{\mathbf{b}_1}+N_1,\cdots,X_{L}^{\mathbf{b}_L}+N_L),U_1^{2i-2};U_{2i})\nonumber\\
=&I((X_{1}^{\mathbf{b}_1\backslash\chi_1^{2i-2}}+N_1,\cdots,X_{L}^{\mathbf{b}_N\backslash\chi_1^{2i-2}}+N_L);U_{2i}),
\end{align}
where $\chi_i^j = \{i, \dots, j\}$ denotes the indicator set. $\mathbf{b}_1 \backslash \chi_1^{2i-2}$ represents the set of bits from the first bit to the $(2i-2)$-th bit, which are known. Based on the polarization unit, $U_{2i}$ is more strongly related to $X_1^{\mathbf{b}_1 \backslash \chi_1^{2i-2}}$ than to $U_{2i-1}$. Similarly, we consider $I(Y_1^4, U_1^2; U_3)$ as an example,
\begin{align}
I(Y_1^4,U_1^{2};U_{3})=&I(X_1^4+N_1^4,U_1^{2};U_{3})\nonumber\\
=&I(\breve{X}_1^4+N_1^4;U_{3}),
\end{align}
where $X_1^4=(U_1\oplus U_2\oplus U_3\oplus U_4,U_3\oplus U_4,U_2\oplus U_4,U_4)$ and $\breve{X}_1^4=(U_3\oplus U_4,U_3\oplus U_4,U_4,U_4)$. It is evident that $I(Y_1^4,U_1^{2};U_{4})\geq I(Y_1^4,U_1^{2};U_{3})$. A similar procedure can be applied to obtain $I(Y_1^L,U_1^{2i-2};U_{2i})\geq I(Y_1^L,U_1^{2i-2};U_{2i-1})$, leading to the conclusion that $I(W_{L}^{(2i)})\geq I(W_{L}^{(2i-1)})$. This analysis can also be used to understand the partial order (PO) of polar codes.

\textbf{Remark 6:} It can also be shown that $I(W_{2L}^{(2i)}) + I(W_{2L}^{(2i-1)}) \geq 2I(W_L^{(i)})$. To avoid redundancy, we provide the proof in appendix \ref{appendix_2}. This indicates that the sum of the mutual information increases after each polarization unit, and the equality is achieved for the memoryless noise.

Finally, we present a representative example to illustrate the increase in mutual information resulting from polarization. We consider the student distribution due to its generality, as it reduces to the Cauchy distribution when $\nu = 1$ and approaches the Gaussian distribution as $\nu \to +\infty$ where $\nu$ denotes the degree of freedom (DoF). For simplicity and tractability, we focus on the case of a single-layer polarization, since the computational complexity of mutual information evaluation increases rapidly with the dimension.
\begin{example}\label{example_1}
\textit{Let $(N_i, N_{i+1})$ follow a two-dimensional student distribution with correlation matrix $\Sigma$ and DoF $\nu$. We set $\Sigma(1,:) = [1, 0.6]$ and $\nu = 2$. The corresponding mutual information comparison is illustrated in Fig.~\ref{fig_3}.}
\vspace{-0cm}
\begin{figure}[htbp]
\centering
\includegraphics[width=4in]{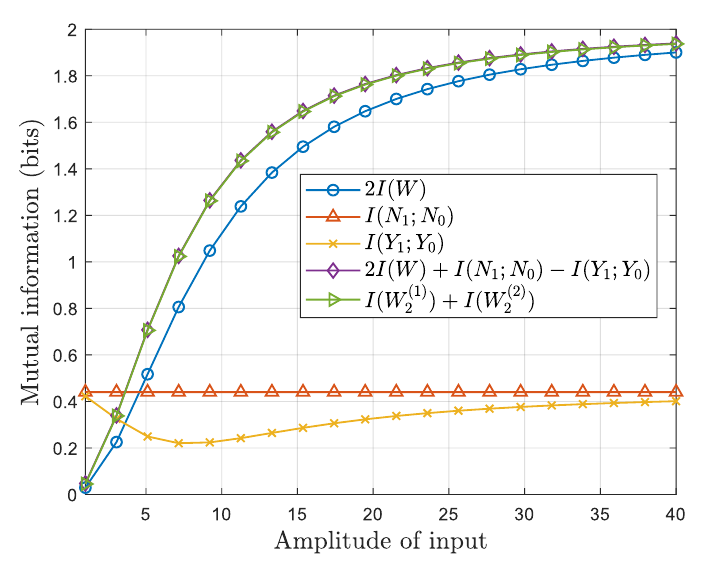}\label{polarization_verification}
\caption{The MI comparison after the single-layer polarization. The considered input amplitude range is $[1,40]$.}
\label{fig_3}
\end{figure}
\end{example}

From Fig. \ref{fig_3}, it can be seen that $I(W_2^{(1)}) + I(W_2^{(2)}) > 2I(W)$ with the additional mutual information quantified by $I(N_0; N_1) - I(Y_0; Y_1)$, which corroborates the preceding theoretical analysis. The term $I(N_0; N_1)$ depends solely on the noise PDF, whose parameters remain fixed in Fig. \ref{fig_3}. Notably, $I(Y_1; Y_2)$ is not a monotonic function of the transmitted signal power. A similar phenomenon and its underlying explanation can be found in our previous work \cite{paper26}.

\section{Characteristics of $\{Z_n\}$}\label{PROOF_OF_THEOREM_4_5}
\subsection{Convergence of $\{Z_n\}$}
Due to the assumption of first-order noise memory, we first consider the transformation of the Bhattacharyya parameter from $L=1$ to $L=2$. We then discuss the convergence of the process $\{Z(W_L^{(i)})\}$ and show that $Z(W_L^{(i)}) \to 0 \text{ or } 1 \text{ a.s.}$ as $L \to +\infty$. The differences between $Z(W)$ and $Z_g(W)$ are also illustrated via examples.

From \cite{paper1}, the mutual information can be bounded by the Bhattacharyya parameter as follows,
\begin{equation}\label{bound_of_Z_I_1}
I(W)\geq\log\frac{2}{1+Z(W)},
\end{equation}
\begin{equation}\label{bound_of_Z_I_2}
I(W)^2+Z(W)^2\leq1.
\end{equation}

Obviously, we have $Z_g(W)\leq1$. First, we consider the case $L=2$, and $Z(W_2^{(1)})$ can be expressed as in \eqref{theorem_2_long_equation_1}. By applying similar manipulations, we can obtain
\begin{align}\label{Z_N_2}
Z(W_2^{(1)})=&Z(W_{a})\geq Z_g(W),\\
Z(W_2^{(2)})=&Z(W_{b})\leq Z_g(W),
\end{align}
where
\begin{align}
Z(W_{a})=&\frac{1}{2}\sum_{U_2\in\mathcal{B}}\int_{\mathcal{R}^2}\sqrt{W(y_1^2|u_2,u_2)W(y_1^2|u_2,u_2\oplus1)}dy_1^2,\nonumber\\
Z(W_{b})=&\frac{1}{2}\sum_{U_1\in\mathcal{B}}\int_{\mathcal{R}^2}\sqrt{W(y_1^2|0,u_1)W(y_1^2|1,u_1\oplus1)}dy_1^2.\nonumber
\end{align}

$Z(W_{b})$ can be interpreted as a symbol being transmitted twice over the same channel. From the perspective of detection based on the Euclidean distance, $Z(W_{a})$ and $Z(W_{b})$ separately corresponds to $\sqrt{2}d$ and $d$, where $d$ is the distance between two adjacent constellation points. In fact, the transformation from $Z(W)$ to $Z(W_2^{(1)})$ and $Z(W_2^{(2)})$ does not affect the entire process $\{Z_n\}$. In other words, for first-order noise memory, the convergence behavior of $\{Z_n\}$ can be analyzed by taking $L=2$ as the starting point and initial state.

Moreover, it is important to note that the relationship between $Z(W_2^{(1)})$ and $Z(W)$ is generally uncertain and cannot be explicitly characterized. We will illustrate this phenomenon via an example in the following. In the next theorem, we investigate the convergence of $\{Z_n\}$ as $L$ goes to infinity.

\begin{figure*}[tb]
\hrulefill
\begin{align}\label{theorem_2_long_equation_1}
Z(W_2^{(1)})=&\frac{1}{2}\int_{\mathcal{R}^2}\sqrt{\mathbb{E}_{N_0}\left(\sum_{U_2}\tilde{W}(y_1|u_2\oplus0,n_0)\tilde{W}(y_2|u_2,u_2\oplus0,y_1)\right)\mathbb{E}_{N_0}\left(\sum_{U_2}\tilde{W}(y_1|u_2\oplus1,n_0)\tilde{W}(y_2|u_2,u_2\oplus1,y_1)\right)}dy_1^2\nonumber\\
=&\frac{1}{2}\int_{\mathcal{R}^2}\sqrt{\sum_{U_2}W(y_1^2|u_2,u_2\oplus0)\sum_{U_2}W(y_1^2|u_2,u_2\oplus0)}dy_1^2.
\end{align}
\end{figure*}

\begin{theorem}\label{theorem_4}
\textit{For the block length $L>2$,
\begin{align}\label{Z_N_larger_than_4_1}
Z(W_{2L}^{(2i-1)})\leq& Z(W_{L}^{(i)})+Z_g(W_{L}^{(i)})-Z(W_{L}^{(i)}){\inf}_{y_L}Z_g(\tilde{W}_{L}^{(i)}|y_L),
\end{align}
\begin{equation}\label{Z_N_larger_than_4_2}
Z(W_{2L}^{(2i-1)})\geq\frac{1}{2}\big(Z(W_{L}^{(i)})+Z_g(W_{L}^{(i)})\big),
\end{equation}
\begin{equation}\label{Z_N_larger_than_4_3}
Z(W_{2L}^{(2i)})\leq Z(W_{L}^{(i)}){\sup}_{y_L}Z_g(\tilde{W}_{L}^{(i)}|y_L),
\end{equation}
\begin{equation}\label{Z_N_larger_than_4_4}
Z(W_{2L}^{(2i-1)})+Z(W_{2L}^{(2i)})\leq Z(W_{L}^{(i)})+Z_g(W_{L}^{(i)})\leq2Z(W_{L}^{(i)}).
\end{equation}}
\end{theorem}
\begin{IEEEproof}
The proof is relegated to appendix \ref{appendix_3}.
\end{IEEEproof}

Subsequently, it can be shown that the sequence $\{Z(W_L^{(i)})\}$ forms a martingale process and $Z(W_L^{(i)})\rightarrow0\text{ or }1\text{ a.s.}$ with respect to $L$. Note that  \eqref{bound_of_Z_I_1} and \eqref{bound_of_Z_I_2} are still valid under our parameter definitions. Then, based on them, we have the following corollary.

\begin{corollary}\label{corollary_5}
\textit{Let the $\{I(W_L^{(i)})\}$ be a sequence with respect to $L$. Then, $I(W_L^{(i)})$ converges to 0 or 1 almost surely as $L\rightarrow+\infty$ for any fixed $i$.}
\end{corollary}

Moreover, the proportion of noiseless subchannels is characterized in the following corollary.
\begin{corollary}\label{corollary_1}
\textit{For any $\epsilon>0$, we have
\begin{align}
\lim\limits_{L\rightarrow+\infty}\frac{1}{L}|\{i:I(W_L^{(i)})>1-\epsilon\}|=I(W)+I^{\dagger}(W).
\end{align}}
\end{corollary}

From \eqref{Z_N_larger_than_4_2} and \eqref{Z_N_larger_than_4_4}, we observe that the decay rate of the Bhattacharyya parameter under channel noise with memory is slower than that for memoryless channels. However, the convergence of $\{Z_n\}$ indicates that polarization still occurs in channels with memory. Additionally, the transformation of $Z(W_L^{(i)})$ is similar to the relation for parallel channel polarization \cite{paper14}. Nevertheless, the underlying reasons why the square transformation for the `good' subchannel does not hold are different. In parallel polarization, the underlying reason is that the original channels may not be identical, whereas theorem \ref{theorem_4} arises due to the memory in channel noise.

We now present an important observation regarding $Z_g(\tilde{W}|n_0)$. From the definition of $Z_g(\tilde{W}|n_0)$, its value is influenced by the previous noise sample $n_0$. On the other hand, $Z_g(\tilde{W}|n_0)$ becomes small if the variance of the channel noise $n_1$ given $n_0$ is small. In other words, the variance of $n_1$ must be a function of $n_0$. If $n_0$ only affects the expectation of $n_1$, then $Z_g(W)$ should equal $Z(W)$. In this case, the Bhattacharyya parameter evolution and polarization rate would be the same as the case in \cite{paper1}. \textbf{This implies that the memory of the channel noise does not necessarily impact the polarization rate unless the statistical distribution of the channel noise meets certain conditions.} Based on these analyses, the following proposition can be obtained.

\begin{proposition}\label{proposition_2}
\textit{Let the noise memory order be $p$ and define the noise random vector as $\vec{N} = [N_1, N_2, \ldots, N_{p+1}]$. If the variables $N_j$, for $j=1,\ldots,p$, influence only the expectation of the conditional distribution $f(n_{p+1}|n_1^p)$, then we have $Z_g(W) = Z(W)$ and the polarization rate does not degrade in the presence of noise memory.}
\end{proposition}

Proposition \ref{proposition_2} provides a necessary condition under which the polarization rate remains unchanged compared to the memoryless case. This result can be directly verified based on the proof of theorem \ref{theorem_4}. To help clarify the differences, we will provide two examples.
\begin{example}\label{example_2}
\textit{Let the random vector $\vec{V}=[V_0,V_1]$ follow a bivariate Gaussian distribution $\mathcal{G}(\vec{\mu}=\vec{0},\Sigma)$. Denote the diagonal elements of $\Sigma^{-1}$ by $\sigma^{-2}$. The GA channel at the first time instant is given by $\tilde{W}(y_1|x_1,v_0)=f_{V_1|V_0}(y_1-x_1|,v_0)$ where $f_{V_1|V_0}(\cdot|\cdot)$ is the conditional PDF of $V_1$ given $V_0$. Based on the \eqref{definition_Z_g}, it can be calculated that}
\begin{equation}
Z_g(W)=Z(W)=\exp\left(-\frac{8}{\sigma^{2}}\right).
\end{equation}
\end{example}

It can be observed that for colored Gaussian noise, the process $\{Z(W_L^{(i)})\}$ is identical to $\{Z_g(W_L^{(i)})\}$. This is because $v_0$ only affects the mean of $v_1$ without altering its variance, which is consistent with the previous analysis. As a result, $Z_g(W|v_0)$ remains the same for any given $v_0$.

\begin{figure}[htbp]
\centering
\includegraphics[width=4in]{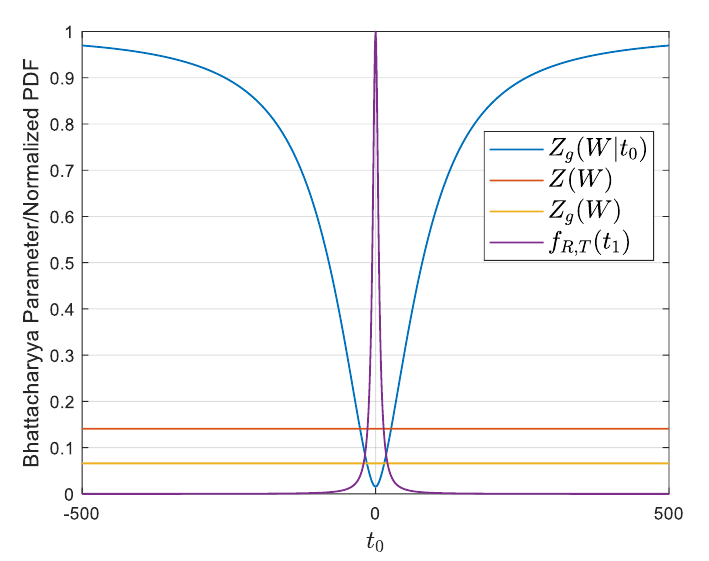}
\caption{Bhattacharyya parameter comparison for channel noise following a bivariate student distribution. To enhance the visibility of $f_T(t_1)$, we plot its normalized version $f_{T,R}(t_1)=f_T(t_1)/\max(f_T(t_1))$. In simulation, we set $\nu=1.2$, $\Sigma(1,:)=(18,12.6)$ and the input signal amplitude $x\in\{0,100\}$. $\Sigma$ is a Topelitz matrix and can be fully constructed by its first row.}
\label{fig_4}
\end{figure}

\begin{example}\label{example_3}
\textit{Let the random vector $\vec{T} = [T_0, T_1]$ follow a bivariate student distribution $\mathcal{T}(\vec{\mu} = \vec{0}, \Sigma, \nu)$. The GA channel at the first time instant is given by $\tilde{W}(y_1|x_1,t_0)=f_{T_1|T_0}(y_1-x_1|,t_0)$ where $f_{T_1|T_0}(\cdot|\cdot)$ is the conditional PDF of $T_1$ given $T_0$. Denote
\begin{equation}
\Sigma=
\left(
\begin{array}{cc}
  \Sigma_{1,1} & \Sigma_{1,2} \\
  \Sigma_{2,1} & \Sigma_{2,2}
\end{array}
\right).
\end{equation}}

\textit{Then, the $f_{T_1|T_0}(t_1|t_0)$ can be obtained as}
\begin{align}\label{conditional_student_PDF}
f_{T_1|T_0}(t_1|t_0)=\frac{\Gamma(\frac{\nu+2}{2})}{\Gamma(\frac{\nu}{2})\nu\pi\sqrt{\det(\Sigma)}}\bigg(1+\frac{\frac{t_0^2}{\Sigma_{1,1}}+\frac{(t_1-\delta)^2}{\sigma}}{\alpha}\bigg)^\frac{-\alpha-p}{2},
\end{align}
\textit{where $f_{T}(\cdot)$ is the marginal PDF and $\delta=t_0\Sigma_{1,2}/\Sigma_{1,1}$, $\sigma=\det(\Sigma)/\Sigma_{1,1}$. The $Z_g(W)$ and $Z(W)$ cannot be computed analytically and thus, we provide a numerical comparison in Fig. \ref{fig_4}.
}
\end{example}

According to Fig. \ref{fig_4}, it can be observed that $Z_g(\tilde{W}|t_0)$ reaches its minimum when $t_0 = 0$, where $\Sigma_{1,1}^{-1}t_0^2$ in \eqref{conditional_student_PDF} becomes zero. Note that a smaller value of $\Sigma_{1,1}^{-1}t_0^2$ results in a more concentrated distribution, which can be interpreted as the PDF having a smaller effective variance although $\Sigma_{1,1}^{-1}t_0^2$ is not the actual variance. Therefore, $Z_g(\tilde{W}|t_0)$ increases monotonically as $|t_0|$ increases. The quantity $Z_g(W)$ is the weighted sum of $Z_g(\tilde{W}|t_0)$ over the marginal distribution $f_T(t_1)$. It is evident that $Z_g(W)$ is strictly smaller than $Z(W)$, which aligns with the theoretical analysis presented earlier.

\subsection{Polarization rate}
The polarization rate is a crucial metric directly related to the performance of polar codes. A higher polarization rate indicates more effective polarization for a fixed block length, which is also associated with a smaller rate loss. Subsequently, we first introduce a lemma that will be used in the following analysis.
\begin{lemma}[\cite{paper10}]\label{lemma_4}
\textit{Let $x$ be a constant. Define the vector $\mathbf{b}=(b_n,b_{n-1},\cdots,b_1)\in\mathcal{B}^n$ and}
\begin{equation}
x_{j+1}(\mathbf{b})=\left\{
\begin{aligned}
&2x_j(\mathbf{b}),b_{j}=1 \\
&x_j(\mathbf{b})+1,b_{j}=0. \\
\end{aligned}
\right.
\end{equation}

\textit{Then, for any fixed $n$,
\begin{equation}
\min\limits_{\mathbf{b}\in\mathcal{B}^n}x_{n+1}(\mathbf{b})=2^{H(\mathbf{b})}+n-H(\mathbf{b}),
\end{equation}
where $H(\cdot)$ represents the Hamming weight.}
\end{lemma}

Lemma \ref{lemma_4} is the dual result of that presented in \cite{paper10}, where $\max\limits_{\mathbf{b}\in\mathcal{B}^n}x_{n+1}(\mathbf{b})$ is derived. Since the proof follows the same steps as in \cite{paper10}, it is omitted here. With the aid of lemma \ref{lemma_4}, the following theorem is obtained.

\begin{theorem}\label{theorem_5}
\textit{Let the block length be $L$ and $i\in\{1,2,\cdots,L\}$. Then,
\begin{equation}
\lim\limits_{L\rightarrow+\infty}\Pr[Z(W_L^{(i)})\leq c_0L^{\frac{c_1}{2}}]=I(W)+I^{\dagger}(W),
\end{equation}
\begin{equation}
\lim\limits_{L\rightarrow+\infty}\Pr[Z(W_L^{(i)})\geq 2^{c_2\sqrt{L}+c_3}]=1,
\end{equation}
where $c_0>0$, $c_1<0$, $c_2<0$ and $c_3$ are some constants.}
\end{theorem}
\begin{IEEEproof}
The proof is relegated to appendix \ref{appendix_4}.
\end{IEEEproof}

\section{Extensions and Future directions}\label{EXPANSION_AND_DISCUSSION}
The previous analysis mainly focuses on the case with noise memory order equal to 1 and uniform binary input. Recall that the channel noise memory order is $p$. In this section, we discuss the extensions to more general cases and provide some future work directions.

\subsection{General cases}
\subsubsection{Memory order $p>1$}
Assume that the channel noise process forms the $p$-th Markov process and the PDF of the channel noise is $f_N(n_1|n_{1-p}^0)$. In this case, the PDF of the underlying channel without prior information and GA channel are defined as $_{p}W:\mathcal{X}\rightarrow\mathcal{Y}$ and $_{p}\tilde{W}:\mathcal{X}^{p+1}\times\mathcal{Y}^{p}\rightarrow\mathcal{Y}$. In Section \ref{PROOF_OF_THEOREM_1}, the key aspect of the proof involves selecting variables for the mutual information in such a way that the mutual information of high-level polarized channels can be decoupled from that of low-level polarized channels, as shown in \eqref{lemma_1_equation_2} and \eqref{split_temp_I_1}. It can be verified that this construction remains valid for $p>1$ and a similar conclusion can be obtained. For example, we present the modified version of theorem \ref{theorem_1} with $p>1$.
\begin{theorem}
\textit{Let the noise memory order $p>1$ and $L$ be the block length. Then,
\begin{align}
\lim\limits_{L\rightarrow+\infty}\frac{1}{L}\sum_{i=1}^{L}I(_{p}W_L^{(i)})=I(_{p}W)+I^{\dagger}(_{p}W),
\end{align}
where
\begin{align}
I^{\dagger}(_{p}W)=I(N_1;N^0_{1-p})-\frac{1}{2}I(Y_1;Y_0)-I(Y_0;Y_1^{+\infty}).
\end{align}}
\end{theorem}

The difference between $I^{\dagger}(_{p}W)$ and $I^{\dagger}(W)$ is merely the constant $I(N_1;N^{-1}_{1-p}|N_0)$. Additionally, $I^{\dagger}(_{p}W)$ is an increasing function with respect to $p$. This implies that the polarization operation extracts the channel noise relation more effectively as $p$ becomes larger. In fact, without the application of polar codes, it is challenging to directly exploit the noise correlation to improve communication performance. It is clear that $I^{\dagger}(_{p}W)$ should be less than or equal to $1-I(_{p}W)$ since $I(_{p}W)+I^{\dagger}(_{p}W)$ represents the ratio of perfect channel. Therefore, the $I(N_1;N^0_{1-p})$ must converge to a constant as $p\rightarrow+\infty$.

\subsubsection{$q$-ary polarization ($q>2$)}
$q$-ary input has been explored to design high-dimensional polarization kernels and improve the error exponent, which is associated with the asymptotic decoding error probability. The main challenge arises from the fact that the relationship between the Bhattacharyya parameter and mutual information for the binary input scheme is not tight enough for $q$-ary input. In \cite{paper15}, Park et al. propose the average Bhattacharyya parameter and more general inequalities between $I(W)$ and $Z(W)$. Using this approach, polarization can still be proved through the martingale-based method, as was employed in the previous analysis. Furthermore, it can be observed that the binary input assumption is not necessary for the main proof. Consequently, the previous discussion and conclusions can be naturally extended to the $q$-ary input alphabet.

\subsubsection{Non-uniform input distribution}
In addition to the extension from binary input to $q$-ary input, the input distribution can also be arbitrary, such as non-uniform distributions. For example, consider $U_j\in\mathsf{GF}(2)$ where $U_j$ follows a Bernoulli distribution, i.e., $\Pr[U_j=1]=1-\Pr[U_j=0]=a$. Let $X_1^2=U_1^2F$ and it can be checked that the $X_1$ is not independent of $X_2$ when $a\neq\frac{1}{2}$. In this case, the proof of theorem \ref{theorem_1} is not valid because the construction of $I_{i}^{j,(k)}(\tilde{W})$ is based on the assumption that all $V_i^{(j)}$ are mutually independent, which holds only when $a=\frac{1}{2}$. In fact, the case where $a\neq\frac{1}{2}$ is more akin to the scenario where the source has memory. However, it is not equivalent to that the input does not form a Markov process and therefore, the analysis would become much more complicated, which is beyond the scope of this paper. In this case, the Honda-Yamamoto scheme should be considered which is designed for polar codes with asymmetric input distributions \cite{paper17}. In fact, extending the analysis to the case of nonuniform input distributions remains an important direction for future work. Moreover, in many scenarios, the channel capacity is achieved with symmetric input. Therefore, we omit the corresponding proofs.

\subsection{Future research directions}
\subsubsection{Rate loss with finite block length}
The gap between the MI without prior information and the achievable capacity was derived previously. However, in practical scenarios, the block length is finite, resulting in incomplete polarization and a consequent rate loss. Unfortunately, the cutoff rate for channels with noise memory is considerably more complex than that for channel with independent noise, and the simple explicit relationship to the Bhattacharyya parameter given in \cite{paper11} no longer holds. This relationship is detailed as follows,
\begin{align}
R(W)=-\log\left(\frac{1}{2}+\frac{1}{2}Z(W)\right).
\end{align}

Based on \cite{paper13}, without considering the prior noise information, it can be generally deduced that
\begin{align}\label{cutoff_rate_for_memory}
R(W,\rho)=\lim\limits_{L\rightarrow+\infty}-\frac{1}{L}\ln\int_{y_1^L\in\mathcal{R}^L}\left(\sum_{x_1^L\in\mathcal{B}^L}p(x_j)\left(W(y_1|x_1)\prod_{j=2}^{L}W(y_j|x_j,y_{j-1},x_{j-1})\right)^{\frac{1}{\rho+1}}\right)^{\rho+1}dy_1^L,
\end{align}
where the base of the logarithm operation `$\ln$' is the natural constant. From \eqref{cutoff_rate_for_memory}, establishing a general relationship between $R(W)$ and $Z(W)$ proves to be highly challenging. Therefore, the asymptotic properties of the cutoff rate process $\{R(W_L^{(i)})\}$ with respect to $L$ may need to be studied independently. An alternative approach to bounding the gap between capacity and cutoff rate is through the estimation of the error exponent. For instance, \cite{paper25} derives the error exponent as a function of block length and the difference between capacity and achievable rate for BMCs. Whether this approach and its conclusions can be extended to channels with memory remains an interesting problem.

\subsubsection{Faster polarization}
Based on theorem \ref{theorem_4}, the polarization rate significantly slows down if the channel noise model satisfies the condition in proposition \ref{proposition_2}. Consequently, achieving the same coding performance as in BMC scenarios requires a larger block length, leading to increased decoding complexity or performance degradation. Therefore, accelerating the polarization process via novel polarization schemes, generalized polarization kernels, channel model transformation, or other possible approaches should be a highly promising research direction.

\section{Conclusion}\label{CONCLUSION}
In this paper, we demonstrated that channel polarization occurs under noise with memory. Particularly, with a sufficient polarization, the average capacity of the subchannels was achievable and larger than that of the underlying channel without considering memory effects. This improvement arises from the utilization of noise correlation during the polarization process, indicating that polar codes are especially well-suited for channels with memory. We further analyzed the polarization rate in terms of the Bhattacharyya parameter and showed that it is slower compared to the memoryless case. These results imply that the performance degradation caused by incomplete or missing prior noise information can be asymptotically eliminated as the block length increases.

\begin{appendices}
\section{Proof of theorem \ref{theorem_1}}\label{appendix_1}
\setcounter{equation}{0}
\renewcommand{\theequation}{A.\arabic{equation}}
In \cite{paper1}, the quantities $I(W_{2L}^{(2i-1)})$ and $I(W_{2L}^{(2i)})$ are expressed in terms of $I(W_{L}^{(i)})$, a representation that naturally extends from the simplest case of $L=2$. However, this extension is not straightforward in our scenario. For instance, one can calculate that
\begin{align}
I(\tilde{W}_{4}^{(1)})+I(\tilde{W}_{4}^{(2)})=&2I(\tilde{W}_{2}^{(1)})+I(N_0;Y_1^4,U_1)\nonumber\\
&-I(Y_3^4;X_2|U_2,Y_2)-I(Y_1^2;Y_3^4)+I(Y_3^4;U_1\oplus U_2,Y_1|U_2,Y_2,N_0).
\end{align}

The RHS of the equation is too complex to simplify due to the noise memory. To handle this problem, we define that
\begin{align}\label{temp_I}
I_{i}^{j,(k)}(\tilde{W})=I(Y_i^j;V_i^{j,(k)},N_0).
\end{align}

\begin{figure}[htbp]
\centering
\includegraphics[width=3in]{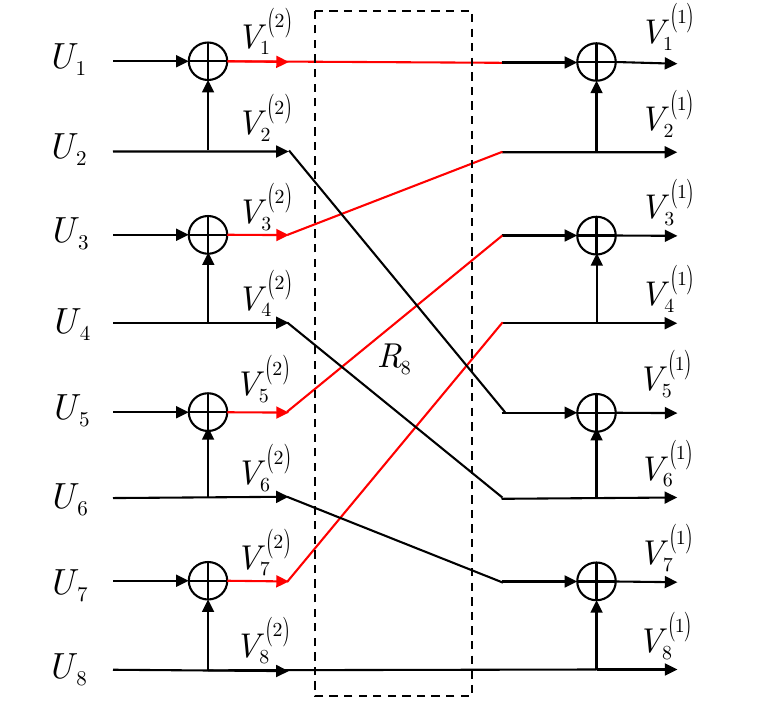}
\caption{The input set of $V_1^{4,(2)}$ with block length $L=8$.}
\label{fig_2}
\end{figure}

The $I_{1}^{L,(l-1)}$ can be separated as
\begin{align}\label{split_temp_I_1}
I_{1}^{L,(l-1)}(\tilde{W})=&I(Y_1^{L/2};V_1^{L,(l-1)},N_0)+I(Y_{L/2+1}^N;V_1^{L,(l-1)},N_0|Y_1^{L/2})\nonumber\\
=&I(Y_1^{L/2};V_{\pi_L^{-1}(\chi_{L/2+1}^{L})}^{(l-1)}|V_{\pi_L^{-1}(\chi_{1}^{L/2})}^{(l-1)},N_0)+I(Y_1^{L/2};V_{\pi_L^{-1}(\chi_{1}^{L/2})}^{(l-1)},N_0)\nonumber\\
&-I(Y_{L/2+1}^L;Y_1^{L/2})+I(Y_{L/2+1}^L;V_1^{L,(l-1)},Y_1^{L/2},N_0),
\end{align}
where $V_i^{j,(k)}$ denotes the inner bits of the polarization block, as illustrated in Fig. \ref{fig_1}. The notation $\pi_L(\cdot)$ represents the index permutation corresponding to $\mathcal{R}_L$, while $\pi_L^{-1}$ refers to the inverse operation. For example, let $L=8$ and then, $V_{\pi_8^{-1}(\chi_1^{4})}^{(2)}$ is equivalent to $V_{\{1,3,5,7\}}^{(2)}$ which is shown in the Fig. \ref{fig_2} via the red line. Note that $I(Y_1^{L/2};V_{\pi_L^{-1}(\chi_{1}^{L/2})}^{(l-1)},N_0)=I_{1}^{L/2,(l-2)}(\tilde{W})$ and $I(Y_1^{L/2};V_{\pi_L^{-1}(\chi_{L/2+1}^{L})}^{(l-1)}|V_{\pi_L^{-1}(\chi_{1}^{L/2})}^{(l-1)},N_0)=0$ since $Y_1^{L/2}$ is only related to $V_{\pi_L^{-1}(\chi_{1}^{L/2})}^{(l-1)}$ and $N_0$. Furthermore,
\begin{align}\label{split_temp_I_2}
I(Y_{L/2+1}^L;V_1^{L,(l-1)},Y_1^{L/2},N_0)=&I(Y_{L/2+1}^L;V_{L/2+1}^{L,(l-1)},Y_{L/2})+I(Y_{L/2+1}^L;V_{1}^{L/2,(l-1)},N_0|V_{L/2+1}^{L,(l-1)},Y_{L/2})\nonumber\\
\overset{(a)}{=}&I(Y_{L/2+1}^L;V_{L/2+1}^{L,(l-1)},N_{L/2})\nonumber\\
\overset{(b)}{=}&I_{1}^{L/2,(l-2)}(\tilde{W}),
\end{align}
where $(a)$ holds because the input of $Y_{L/2}$ is $V_{L-1}^{(l-1)}$, which belongs to $V_{L/2+1}^{L,(l-1)}$. Here, we clarify that $V_{L/2+1}^{L,(l-1)}$ is equivalent to $V_{\{L/2+1,\cdots,L\}}^{(l-1)}$. Additionally, $Y_{L/2+1}^L$ is independent of $V_{1}^{L/2,(l-1)}$ and $N_0$ given $V_{L/2+1}^{L,(l-1)}$ and $Y_{L/2}$. $(b)$ follows equation \eqref{temp_I}. By plugging \eqref{split_temp_I_2} into \eqref{split_temp_I_1}, we get
\begin{align}\label{recursive_relation_temp_I}
I_{1}^{L,(l-1)}(\tilde{W})=2I_{1}^{L/2,(l-2)}(\tilde{W})-I(Y_{L/2+1}^L;Y_1^{L/2}).
\end{align}

Based on the recursive equation, it can be deduced that
\begin{align}\label{split_temp_I_3}
I_{1}^{L,(l-1)}(\tilde{W})=&\frac{L}{2}I(Y_1^2;X_1^2,N_0)-\sum_{i=1}^{\log L}2^{i-1}I(Y_1^{2^{\log L-i}};Y_{2^{\log L-i}+1}^{2^{\log L-i+1}})\nonumber\\
\overset{(a)}{=}&\frac{L}{2}I(Y_1^2;X_1^2,N_0)-\sum_{i=1}^{L-1}I(Y_0;Y_1^i)\nonumber\\
\overset{(b)}{=}&LI(\tilde{W})-\frac{L}{2}I(Y_0;Y_1)-\sum_{i=1}^{L-1}I(Y_0;Y_1^i),
\end{align}
where $(a)$ is due to the lemma \ref{lemma_3} and $(b)$ is based on the \eqref{lemma_1_equation_2}. Then,
\begin{align}
\frac{1}{L}\sum_{i=1}^{L}I(\tilde{W}_L^{(i)})=&\frac{1}{L}\sum_{i=1}^{L}I(Y_1^L;U_i,N_0|U_1^{i-1})\nonumber\\
=&\frac{1}{L}\left(I(Y_1^L;U_1,N_0)+\sum_{i=2}^{L}I(Y_1^L;U_i|N_0,U_1^{i-1})+\sum_{i=2}^{L}I(Y_1^L;N_0|U_1^{i-1})\right)\nonumber\\
=&\frac{1}{L}I(Y_1^L;U_1^L,N_0)+\frac{1}{L}\sum_{i=2}^{L}I(Y_1^L,U_1^{i-1};N_0).
\end{align}

Note that $I_{1}^{L,(l-1)}(\tilde{W})=I(Y_1^L;U_1^L,N_0)$ since there is a bijection between $U_1^L$ and $V_1^{L,(l-1)}$. In accordance with \eqref{split_temp_I_3},
\begin{align}\label{LHS_of_theorem_1_1}
\frac{1}{L}\sum_{i=1}^{L}I(\tilde{W}_L^{(i)})=I(\tilde{W})-\frac{1}{2}I(Y_0;Y_1)-\frac{1}{L}\sum_{i=1}^{L-1}I(Y_0;Y_1^i)+\underbrace{\frac{1}{L}\sum_{i=2}^{L}I(Y_1^L,U_1^{i-1};N_0)}_{\triangleq \beta(L)}.
\end{align}

From the corollary \ref{corollary_4}, \eqref{LHS_of_theorem_1_1} can be simplified as
\begin{align}
\lim\limits_{L\rightarrow+\infty}\frac{1}{L}\sum_{i=1}^{L}I(\tilde{W}_L^{(i)})=&I(\tilde{W})-\frac{1}{2}I(Y_0;Y_1)-I(Y_0;Y_1^{+\infty})+\lim\limits_{L\rightarrow+\infty}\beta(L).
\end{align}

The $\tilde{W}$ should be transformed to the $W$ by $I(\tilde{W}_L^{(i)})=I(W_L^{(i)})+I(Y_1^L;N_0|U_1^{i})$ and $I(\tilde{W})=I(W)+I(N_0;N_1)$. Consequently,
\begin{align}
\lim\limits_{L\rightarrow+\infty}\frac{1}{L}\sum_{i=1}^{L}I(W_L^{(i)})=&I(W)+I(N_0;N_1)-\frac{1}{2}I(Y_0;Y_1)-I(Y_0;Y_1^{+\infty})\nonumber\\
\triangleq& I(W)+I^{\dagger}(W).
\end{align}

Finally, we demonstrate that $I^{\dagger}(W) > 0$, which is sufficient to prove that the sequence $\{\frac{1}{L}\sum_{i=1}^{L}I(W_L^{(i)})\}$ is increasing with respect to $L$. Specifically, we have $\frac{1}{L}\sum_{i=1}^{L}I(W_L^{(i)})=\frac{1}{L}I(Y_1^L;U_1^{L},N_0)-\frac{1}{L}I(N_0;N_1)$ and
\begin{align}
\frac{1}{2L}\sum_{i=1}^{2L}I(W_L^{(i)})-\frac{1}{L}\sum_{i=1}^{L}I(W_L^{(i)})=&\frac{1}{2L}\left(I(Y_1^{2L};U_1^{2L},N_0)-I(N_0;N_1)-2I(Y_1^L;U_1^{L},N_0)+2I(N_0;N_1)\right)\nonumber\\
\overset{(a)}{=}&\frac{1}{2L}\left(I(N_0;N_1)-I(Y_1^{L};Y_{L+1}^{2L})\right)\nonumber\\
=&\frac{1}{2L}\left(I(N_0;N_1)-I(Y_1^{L},X_L;X_{L+1},Y_{L+1}^{2L})\right.\nonumber\\
&\left.+I(Y_1^{L};X_{L+1},Y_{L+1}^{2L}|X_L)+I(Y_1^{L};Y_{L+1}^{2L}|X_{L+1})\right)\nonumber\\
\overset{(b)}{=}&\frac{1}{2L}\left(I(Y_1^{L};X_{L+1},Y_{L+1}^{2L}|X_L)+I(Y_1^{L};Y_{L+1}^{2L}|X_{L+1})\right)\geq0,
\end{align}
where $(a)$ is based on \eqref{recursive_relation_temp_I}. $(b)$ stems from $I(Y_1^{N},X_L;X_{L+1},Y_{L+1}^{2L})=I(N_L;N_{L+1})=I(N_0;N_1)$. Then, we can complete the proof of the theorem \ref{theorem_1}.

\section{Proof of Remark 4}\label{appendix_2}
\setcounter{equation}{0}
\renewcommand{\theequation}{B.\arabic{equation}}
Consider the MI transformation from block length $L$ to $2L$,
\begin{align}\label{appendix_A_1}
I(W_{2L}^{(2j-1)})+I(W_{2L}^{(2j)})=&I(Y_1^{2L};U_{2j-1}|U_1^{(2j-2)})+I(Y_1^{2L};U_{2j}|U_1^{(2j-1)})\nonumber\\
=&I(Y_1^{2L};V_1^{2j,(l)})-I(Y_1^{2L};V_1^{2j-2,(l)}).
\end{align}

Meanwhile, we have $I(W_{L}^{(j)})=I(Y_1^L,V_{1,o}^{2j-2,(l)};V_{2j-1}^{(l)})$. Thus, the \eqref{appendix_A_1} needs to be further decomposed. With some manipulations,
\begin{align}\label{appendix_A_2}
I(Y_1^{2L};V_1^{2j,(l)})=&2I(Y_1^L;V_{1,o}^{2j,(l)})+I(Y_1^L,V_{1,o}^{2j,(l)};V_{1,e}^{2j,(l)})\nonumber\\
&+I(Y_1^L,V_{1,o}^{2j,(l)};Y_{L+1}^{2L},V_{1,e}^{2j,(l)})-I(Y_1^L;Y_{L+1}^{2L}).
\end{align}

The \eqref{appendix_A_2} can be applied for $I(Y_1^{2L};V_1^{2j-2,(l)})$ similarly and therefore,
\begin{align}
I(W_{2L}^{(2j-1)})+I(W_{2L}^{(2j)})=&2I(Y_1^L;V_{1,o}^{2j,(l)})-2I(Y_1^L;V_{1,o}^{2j-2,(l)})+I_r,
\end{align}
where
\begin{align}
I_r=&I(Y_1^L,V_{1,o}^{2j,(l)};V_{1,e}^{2j,(l)})+I(Y_1^L,V_{1,o}^{2j,(l)};Y_{L+1}^{2L},V_{1,e}^{2j,(l)})\nonumber\\
&-I(Y_1^L,V_{1,o}^{2j-2,(l)};V_{1,e}^{2j-2,(l)})-I(Y_1^L,V_{1,o}^{2j-2,(l)};Y_{L+1}^{2L},V_{1,e}^{2j-2,(l)}).
\end{align}

It can be observed that $I(Y_1^L,V_{1,o}^{2j-2,(l)};V_{2j-1}^{(l)})=I(Y_1^L;V_{1,o}^{2j,(l)})-I(Y_1^L;V_{1,o}^{2j-2,(l)})$ and
\begin{align}
I(W_{2L}^{(2j-1)})+I(W_{2L}^{(2j)})=2I(W_{L}^{(j)})+I_r.
\end{align}

Finally, we can verify $I_r\geq0$ by the chain rule and then, we can conclude that $I(W_{2L}^{(2j-1)})+I(W_{2L}^{(2j)})\geq2I(W_{L}^{(j)})$.

\section{Proof of theorem \ref{theorem_4}}\label{appendix_3}
\setcounter{equation}{0}
\renewcommand{\theequation}{C.\arabic{equation}}
\begin{figure*}[tb]
\hrulefill
\begin{align}\label{theorem_2_long_equation_2}
&Z(W_{2L}^{(2i-1)})=\frac{1}{2}\sum_{U_1^{2i-2}}\int_{\mathcal{R}^{2L}}\sqrt{\prod_{\varrho\in\mathcal{B}}\sum_{U_{2i}}\tilde{W}_{L}^{(i)}(y_{L+1}^{2L},u_{1,e}^{2i-2}|u_{2i},n_L)W_{L}^{(i)}(y_{1}^{L},u_{1,e}^{2i-2}\oplus u_{1,o}^{2i-2}|u_{2i}\oplus \varrho)}dy_{1}^{2L}\nonumber\\
\overset{(a)}{\leq}&\frac{1}{2}\sum_{U_1^{2i-2}}\int_{\mathcal{R}^{2L}}\prod_{\varrho\in\mathcal{B}}\sum_{U_{2i}}\sqrt{W_{L}^{e}(u_{2i})W_{L}^{o,e}(u_{2i}\oplus \varrho)}dy_{1}^{2L}-\sum_{U_1^{2i-2}}\int_{\mathcal{R}^{2L}}\sqrt{W_{L}^{e}(0)W_{L}^{o,e}(0)W_{L}^{e}(1)W_{L}^{o,e}(1)}dy_{1}^{2L}\nonumber\\
=&\frac{1}{2}\sum_{U_1^{2i-2}}\int_{\mathcal{R}^{2L}}\sum_{\varrho\in\mathcal{B}}\bigg(\underbrace{W_{L}^{e}(\varrho)\sqrt{W_{L}^{o,e}(0)W_{L}^{o,e}(1)}}_{\triangleq\lambda_{e}(\varrho)}+\underbrace{W_{L}^{o,e}(\varrho)\sqrt{W_{L}^{e}(0)W_{L}^{e}(1)}}_{\triangleq\lambda_{o,e}(\varrho)}\bigg)dy_{1}^{2L}-\sqrt{W_{L}^{e}(0)W_{L}^{o,e}(0)W_{L}^{e}(1)W_{L}^{o,e}(1)}\nonumber\\
=&Z(W_{L}^{(i)})+Z_g(W_{L}^{(i)})-Z(W_{L}^{(i)}){\inf}_{y_L}Z_g(\tilde{W}_{L}^{(i)}|y_L).
\end{align}
\end{figure*}

Based on the definition, $Z(W_{2L}^{(2i-1)})$ can be expressed as \eqref{theorem_2_long_equation_2}, where we define
\begin{equation}
W_{L}^{e}(\varrho)\triangleq \tilde{W}_{L}^{(i)}(y_{L+1}^{2L},u_{1,e}^{2i-2}|\varrho,n_L),
\end{equation}
\begin{equation}
W_{L}^{o,e}(\varrho)\triangleq W_{L}^{(i)}(y_{1}^{L},u_{1,o}^{2i-2}\oplus u_{1,e}^{2i-2}|\varrho).
\end{equation}

The $(a)$ in \eqref{theorem_2_long_equation_2} follows the inequality in appendix D of \cite{paper1}. Note that we will omit the domains of sum arguments for simplicity, provided there is no ambiguity. For $\lambda_{e}(\varrho)$, it can be simplified as
\begin{align}
\sum_{U_1^{2i-2}}\int_{\mathcal{R}^{2L}}\lambda_{e}(\varrho)dy_{1}^{2L}&\overset{(a)}{=}\sum_{U_{1,e}^{2i-2}}\int_{\mathcal{R}^{L}}W_{L}^{e}(\varrho)dy_{L+1}^{2L}\sum_{U_{1,o}^{2i-2}\oplus U_{1,e}^{2i-2}}\int_{\mathcal{R}^{L}}\sqrt{W_{L}^{o,e}(0)W_{L}^{o,e}(1)}dy_{1}^{L}\nonumber\\
&=Z(W_{L}^{(i)}),
\end{align}
where $(a)$ is because that $(U_{1}^{2i-2})$ and $(U_{1,e}^{2i-2},U_{1,o}^{2i-2}\oplus U_{1,e}^{2i-2})$ form a bijection and
\begin{align}
\sum_{U_{1,e}^{2i-2}}\int_{\mathcal{R}^{L}}W_{L}^{e}(\varrho)dy_{L+1}^{2L}=1.
\end{align}

As for $\lambda_{o,e}(\varrho)$, the same simplification does not apply since both $W_{L}^{e}(\varrho)$ and $W_{L}^{o,e}(\varrho)$ are dependent on $y_L$. We can decompose $\lambda_{o,e}(\varrho)$ as follows,
\begin{align}
\sum_{U_1^{2i-2}}\int_{\mathcal{R}^{2L}}\lambda_{o,e}(\varrho)dy_{1}^{2L}=&\sum_{U_{1,o}^{2i-2}\oplus U_{1,e}^{2i-2}}\int_{\mathcal{R}}\int_{\mathcal{R}^{L-1}}W_{L}^{o,e}(\varrho)dy_{1}^{L}\sum_{U_{1,e}^{2i-2}}\int_{\mathcal{R}^{L}}\sqrt{W_{L}^{e}(0)W_{L}^{e}(1)}dy_{L+1}^{2L}\nonumber\\
=&\int_{\mathcal{R}}W_{L}^{(i)}(y_{L}|\varrho)Z(\tilde{W}_{L}^{(i)}|y_L)dy_L=Z_g(W_{L}^{(i)}).
\end{align}

As for the final part of in \eqref{theorem_2_long_equation_2},
\begin{align}
\sum_{U_1^{2i-2}}\int_{\mathcal{R}^{2L}}\sqrt{W_{L}^{e}(0)W_{L}^{o,e}(0)W_{L}^{e}(1)W_{L}^{o,e}(1)}dy_{1}^{2L}\leq&\sum_{U_{1,o}^{2i-2}\oplus U_{1,e}^{2i-2}}\int_{\mathcal{R}^{L}}\sqrt{W_{L}^{o,e}(0)W_{L}^{o,e}(1)}dy_{1}^{L}\nonumber\\
&\times{\sup}_{y_L}\sum_{U_{1,e}^{2i-2}}\int_{\mathcal{R}^{L}}\sqrt{W_{L}^{o,e}(0)W_{L}^{o,e}(1)}dy_{L+1}^{2L}\nonumber\\
=&Z(W_{L}^{(i)}){\sup}_{y_L}Z_g(\tilde{W}_{L}^{(i)}|y_L).
\end{align}

Combining the above analysis, \eqref{Z_N_larger_than_4_1} can be derived. Then,
\begin{align}
Z(W_{2L}^{(2i)})=&\sum_{U_1^{2i-1}}\frac{1}{2}\int_{\mathcal{R}^{2L}}\sqrt{W_{L}^{e}(0)W_{L}^{o,e}(u_{2i-1})}\sqrt{W_{L}^{e}(1)W_{L}^{o,e}(u_{2i-1}\oplus1)}dy_{1}^{2L}\nonumber\\
=&\sum_{U_1^{2i-2}}\int_{\mathcal{R}^{2L}}\sqrt{W_{L}^{e}(0)W_{L}^{o,e}(0)W_{L}^{e}(1)W_{L}^{o,e}(1)}dy_{1}^{2L}\nonumber\\
=&\sum_{U_{1,o}^{2i-2}\oplus U_{1,e}^{2i-2}}\int_{\mathcal{R}}Z_g(\tilde{W}_{L}^{(i)}|y_L)dy_L\int_{\mathcal{R}^{L-1}}\sqrt{W_{L}^{o,e}(0)W_{L}^{o,e}(1)}dy_{1}^{L-1}\nonumber\\
\leq&{\sup}_{y_L}Z_g(\tilde{W}_{L}^{(i)}|y_L)Z(W_{L}^{(i)}).
\end{align}

Based on \eqref{Z_N_larger_than_4_1} and \eqref{Z_N_larger_than_4_3}, we know that $Z(W_{2L}^{(2i-1)})+Z(W_{2L}^{(2i)})\leq2Z(W_{L}^{(i)})+{\sup}_{y_L}Z_g(\tilde{W}_{L}^{(i)}|y_L)-{\inf}_{y_L}Z_g(\tilde{W}_{L}^{(i)}|y_L)$. However, this upper bound is not particularly useful, as it does not guarantee that the sum of the Bhattacharyya parameters of two adjacent split channels decreases as $L$ increases. In fact, a tighter upper bound for $Z(W_{2L}^{(2i-1)}) + Z(W_{2L}^{(2i)})$ can be easily obtained, i.e.,
\begin{align}
Z(W_{2L}^{(2i-1)})+Z(W_{2L}^{(2i)})\leq&Z(W_{L}^{(i)})+Z_g(W_{L}^{(i)})+Z(W_{2L}^{(2i)})\nonumber\\
&-\sum_{U_1^{2i-2}}\int_{\mathcal{R}^{2L}}\sqrt{W_{L}^{e}(0)W_{L}^{o,e}(0)W_{L}^{e}(1)W_{L}^{o,e}(1)}dy_{1}^{2L}\nonumber\\
=&Z(W_{L}^{(i)})+Z_g(W_{L}^{(i)}).
\end{align}

According to the lemma \ref{lemma_1}, we have $Z(W_{2L}^{(2i-1)})+Z(W_{2L}^{(2i)})\leq2Z(W_{L}^{(i)})$. Furthermore, the relation between $Z(W_{2L}^{(2i-1)})$ and $Z(W_{2L}^{(2i)})$ also needs to be determined. Indeed,
\begin{align}
Z(W_{2L}^{(2i-1)})=&\frac{1}{2}\sum_{U_1^{2i-2}}\int_{\mathcal{R}^{2L}}\Big(\sum_{\tau\in\mathcal{B}}W_{L}^{e}(\tau)^2\prod_{\varrho\in\mathcal{B}}W_{L}^{o,e}(\varrho)+\sum_{\tau\in\mathcal{B}}W_{L}^{o,e}(\tau)^2\prod_{\varrho\in\mathcal{B}}W_{L}^{e}(\varrho)\Big)^{\frac{1}{2}}dy_{1}^{2L}\nonumber\\
\geq&\frac{1}{4}\sum_{U_1^{2i-2}}\int_{\mathcal{R}^{2L}}\Big(\sum_{\tau\in\mathcal{B}}\prod_{\varrho\in\mathcal{B}}\sqrt{W_{L}^{e}(\tau)^2W_{L}^{o,e}(\varrho)}+\sum_{\tau\in\mathcal{B}}\prod_{\varrho\in\mathcal{B}}\sqrt{W_{L}^{o,e}(\tau)^2W_{L}^{e}(\varrho)}\Big)dy_{1}^{2L}\nonumber\\
=&\frac{1}{2}(Z(W_{L}^{(i)})+Z_g(W_{L}^{(i)})).
\end{align}

Now, we have that $Z(W_{2L}^{(2i-1)})\geq\frac{1}{2}(Z(W_{L}^{(i)})+Z_g(W_{L}^{(i)}))\geq Z(W_{2L}^{(2i)})$, which completes the proof.

\section{Proof of theorem \ref{theorem_5}}\label{appendix_4}
\setcounter{equation}{0}
\renewcommand{\theequation}{D.\arabic{equation}}
Recall that $l=\log L$. We define that
\begin{equation}
\rho_l^{(i)}\triangleq \frac{Z_g(W_{L}^{(i)})}{Z(W_{L}^{(i)})},
\end{equation}
\begin{equation}
w_l(Y_L^{(i)})\triangleq\sum_{U_{1,o}^{2i-2}\oplus U_{1,e}^{2i-2}}\int_{\mathcal{R}^{L-1}}\sqrt{W_{L}^{o,e}(0)W_{L}^{o,e}(1)}dy_1^{L-1}.
\end{equation}

According to the theorem \ref{theorem_4},
\begin{align}
Z(W_{2L}^{(2i)})=&\int_{\mathcal{R}}Z_g(\tilde{W}_{L}^{(i)}|y_L)w_l(Y_L^{(i)})dy_L\nonumber\\
=&Z(W_{L}^{(i)})\underbrace{\int_{\mathcal{R}}\frac{Z_g(\tilde{W}_{L}^{(i)}|y_L)w_L(Y_L^{(i)})}{Z(W_{L}^{(i)})}dy_L}_{\triangleq\lambda_l^{(i)}}.
\end{align}

Then, we have
\begin{align}
\frac{1+\rho}{2}Z(W_{L}^{^{(i)}})\leq Z(W_{2L}^{(2i-1)})&\leq(1+\rho_l^{(i)}-\lambda_l^{(i)})Z(W_{L}^{^{(i)}}),B_{l}=1,
\end{align}
\begin{align}
Z(W_{L}^{^{(i)}})^2\leq Z(W_{2L}^{(2i)})\leq\lambda_l^{(i)}Z(W_{L}^{^{(i)}}),B_{l}=0.
\end{align}

With the aid of lemma \ref{lemma_4}, we define the lower process $\{S_l\},l\geq2$ and upper process $\{U_l\},l\geq2$ of $\{\log Z(W_L^{(i)})\}$ as follows,
\begin{equation}
S_{l+1}=\left\{
\begin{aligned}
&2S_l\hspace{2.1cm},B_{l}=1 \\
&S_l+\log\frac{1}{2}(1+\rho),B_{l}=0, \\
\end{aligned}
\right.
\end{equation}

\begin{equation}
U_{l+1}=\left\{
\begin{aligned}
&U_l+\log\lambda\hspace{1.4cm},B_{l}=1 \\
&U_l+\log(1+\rho-\lambda),B_{l}=0, \\
\end{aligned}
\right.
\end{equation}
where $\rho\triangleq{\sup}_{i,l}\rho_l^{(i)}$ and $\lambda\triangleq{\inf}_{i,l}\lambda_l^{(i)}$. Without loss of generality, the processes are defined for $l \geq 2$ because the Bhattacharyya parameter transformation from $l = 1$ to $l = 2$ differs from that for $l \geq 2$. We first analyze the upper bound of the polarization rate. For any fixed $\epsilon > 0$ and sufficiently large $l$, we have $\Pr[\sum_{j=2}^{l}B_j>(\frac{1}{2}-\epsilon)(l-1)]>1-\epsilon$. Thus,
\begin{align}\label{theorem_4_temp_1}
U_l=&U_2+(\frac{1}{2}-\epsilon)(l-1)\log\lambda+(\frac{1}{2}+\epsilon)(l-1)\log(1+\rho-\lambda)\nonumber\\
\leq&\frac{l-1}{2}\log\lambda(1+\rho-\lambda)+\epsilon(l-1)\log\frac{1+\rho-\lambda}{\lambda}.
\end{align}

The inequality in \eqref{theorem_4_temp_1} is due to $U_l\leq0$. Hence,
\begin{align}
\lim\limits_{\epsilon\rightarrow0}Z(W_L^{(i)})\leq&2^{U_l}\nonumber\\
\leq&\lim\limits_{\epsilon\rightarrow0}2^{\frac{l-1}{2}\log\lambda(1+\rho-\lambda)+\epsilon(l-1)\log\frac{1+\rho-\lambda}{\lambda}}\nonumber\\
=&2^{-\frac{1}{2}\log\lambda(1+\rho-\lambda)}\sqrt{L}^{\log\lambda(1+\rho-\lambda)}.
\end{align}

For the sake of achieving an arbitrarily small error probability as the block length tends to infinity, we require that $\lambda(1 + \rho - \lambda) < 1$. In fact, it can be verified that this inequality always holds when $\rho \leq 1$. Furthermore, based on \eqref{bound_of_Z_I_1}, \eqref{bound_of_Z_I_2}, and corollary \ref{corollary_1}, we can conclude that
\begin{equation}
\lim\limits_{L\rightarrow+\infty}\Pr[Z(W_L^{(i)})\leq c_0L^{\frac{c_1}{2}}]=I(W)+I^{\dagger}(W),
\end{equation}
where $c_0=2^{-\frac{1}{2}\log\lambda(1+\rho-\lambda)}>0$ and $c_1=\log\lambda(1+\rho-\lambda)<0$. As for the lower bound, the derivation follows a similar process. For instance,
\begin{align}
S_l\geq& 2^{(\frac{1}{2}-\epsilon)(l-1)}S_2+(\frac{1}{2}+\epsilon)(l-1)\log\frac{1}{2}(1+\rho).
\end{align}

The inequality arises from the application of lemma \ref{lemma_4}. Without loss of generality, we set $S_2=\min\limits_{i}\log Z(W_2^{(i)})<0$. After simplifications, the lower bound of $Z(W_L^{(i)})$ can be derived, where $c_2=S_2/\sqrt{2}<0$ and $c_3=(\rho+1)2^{(l-3)/2}>0$.

\end{appendices}

\footnotesize
\bibliographystyle{IEEEtran}
\bibliography{ref}

\end{document}